\documentclass[journal]{IEEEtran}
\usepackage[normalem]{ulem}
\usepackage[noadjust]{cite}
\usepackage{bbm}
\usepackage{array}
\usepackage{dcolumn}
\usepackage{epsfig}
\usepackage[intlimits]{amsmath}
\usepackage{amsmath, amsfonts, yhmath, bm}
\usepackage{amssymb}
\usepackage{amsthm}
\usepackage{psfrag}
\usepackage{color,soul}
\usepackage[dvipsnames]{xcolor}
\usepackage[normalem]{ulem}
\usepackage{enumerate}
\usepackage{stackengine}
\usepackage[noadjust]{cite}
\usepackage{graphicx}
\usepackage{caption}
\usepackage{subcaption}
\usepackage[font=footnotesize]{subcaption}
\usepackage{multirow}
\usepackage{cite}
\usepackage[font=footnotesize]{caption}
\usepackage{etoolbox}
\usepackage{tcolorbox}
\usepackage{float}
\usepackage{dsfont}
\usepackage{tikz}
\usetikzlibrary{arrows}
\newcommand {\myvec}[1] {{\mbox{\boldmath $#1$}}}
\newcommand {\mymat}[1]  {{\mbox{\boldmath $#1$}}}

\newcommand*{\myfontb}{\fontfamily{lmr}\selectfont}

\DeclareMathAlphabet      {\mathbfit}{OML}{cmm}{b}{it}
\DeclareMathAlphabet	  {\mathbfcal}{OMS}{cmsy}{b}{n}

\newcommand{\etal}{\textit{et al.}}

\newcommand {\mS} {\mymat{S}}
\newcommand {\A} {\mymat{A}}

\newcommand {\meps} {\mymat{\mathcal{E}}}

\newcommand {\U} {\mymat{U}}

\newcommand {\bS} {\mybar{\mS}}
\newcommand {\bX} {\mybar{\X}}

\newcommand {\B} {\mymat{B}}
\newcommand {\J} {\mymat{J}}

\newcommand {\hB} {\widehat{\B}}

\newcommand {\C} {\mymat{C}}
\newcommand {\D} {\mymat{D}}

\newcommand {\mGamma} {\mymat{\Gamma}}
\newcommand {\mPhi} {\mymat{\Phi}}

\renewcommand {\H} {\mymat{H}}

\renewcommand {\P} {\mymat{P}}

\newcommand {\Q} {\mymat{Q}}

\newcommand {\R} {\mymat{R}}

\newcommand {\I} {\mymat{I}}
\newcommand {\X} {\mymat{X}}

\newcommand {\ue} {\myvec{e}}

\newcommand {\ub} {\myvec{b}}

\newcommand {\ux} {\myvec{x}}
\newcommand {\ud} {\myvec{d}}

\newcommand {\up} {\myvec{p}}

\newcommand {\uv} {\myvec{v}}

\newcommand {\us} {\myvec{s}}
\newcommand {\usigma} {\myvec{\sigma}}
\newcommand {\uphi} {\myvec{\phi}}

\newcommand {\umu} {\myvec{\mu}}

\newcommand {\uz} {\myvec{z}}
\newcommand {\uh} {\myvec{h}}

\newcommand {\utheta} {\myvec{\theta}}
\newcommand {\hutheta} {\widehat{\myvec{\theta}}}

\newcommand {\Rset} {\mathbb{R}}
\newcommand {\Cset} {\mathbb{C}}

\newcommand {\Eset} {\mathbb{E}}

\DeclareMathOperator{\Tr}{Tr}
\newcommand {\tps} {\mathrm{T}}
\DeclareMathOperator{\Diag}{Diag}
\newcommand {\CRLB} {{\mathrm{CRLB}}}
\newcommand {\MSE} {{\mathrm{MSE}}}

\newcommand\norm[1]{\left\lVert#1\right\rVert}
\newcommand {\sbar} {\mybar{s}}
\newcommand {\xbar} {\mybar{x}}
\newcommand {\vbar} {\mybar{v}}

\newcommand {\buv} {\mybar{\uv}}
\newcommand {\bus} {\mybar{\us}}

\newcommand {\bux} {\mybar{\ux}}
\newcommand {\her} {\mathrm{H}}
\newcommand {\hup} {\widehat{\up}}
\newcommand {\MFP} {\text{\tiny MFP3}}

\newcommand {\SBL} {\text{\tiny SBL}}

\makeatletter
\newsavebox\myboxA
\newsavebox\myboxB
\newlength\mylenA

\newcommand*\mybar[2][0.75]{%
	\sbox{\myboxA}{$\m@th#2$}%
	\setbox\myboxB\null
	\ht\myboxB=\ht\myboxA%
	\dp\myboxB=\dp\myboxA%
	\wd\myboxB=#1\wd\myboxA
	\sbox\myboxB{$\m@th\overline{\copy\myboxB}$}
	\setlength\mylenA{\the\wd\myboxA}
	\addtolength\mylenA{-\the\wd\myboxB}%
	\ifdim\wd\myboxB<\wd\myboxA%
	\rlap{\hskip 0.5\mylenA\usebox\myboxB}{\usebox\myboxA}%
	\else
	\hskip -0.5\mylenA\rlap{\usebox\myboxA}{\hskip 0.5\mylenA\usebox\myboxB}%
	\fi}
\makeatother

\newtheorem{prop}{Proposition}

\DeclareMathOperator*{\argmin}{argmin}
\DeclareMathOperator*{\argmax}{argmax}
\DeclareMathOperator{\rank}{rank}

\def\comment#1{}

\newcommand{\stkout}[1]{
	\color{red}\ifmmode\text{\sout{\ensuremath{#1}}}\else\sout{#1}\fi\color{black}}

\newcommand{\addra}{}
\newcommand{\delra}{\comment}

\begin{document}

\title{A Semi-Blind Method for Localization of Underwater Acoustic Sources}

\author{Amir Weiss, Toros Arikan, Hari Vishnu, Grant B.\ Deane, Andrew C.\ Singer, and Gregory W. Wornell
\thanks{This work was supported, in part, by ONR under Grant Nos.\ N00014-19-1-2661, N00014-19-1-2662, and N00014-19-1-2665, and NSF under Grant No.\ CCF-1816209.}
\thanks{This paper has supplementary material available at http://weissamir.com, provided by the authors. The material includes: (i) technical derivations; (ii) details on the experiment; and (iii) a MATLAB code implementation of the proposed localization method, and computation of the performance bound.}
}

\maketitle

\begin{abstract}
Underwater acoustic localization has traditionally been challenging due to the presence of  unknown environmental structure and dynamic conditions. The problem is richer still when such structure includes occlusion, which causes the loss of line-of-sight (LOS) between the acoustic source and the receivers, on which many of the existing localization algorithms rely. We develop a semi-blind passive localization method capable of accurately estimating the source's position even in the possible absence of LOS between the source and all receivers. Based on typically-available prior knowledge of the water surface and bottom, we derive a closed-form expression for the optimal estimator under a multi-ray propagation model, which is suitable for shallow-water environments and high-frequency signals. By exploiting a computationally efficient form of this estimator, our methodology makes comparatively high-resolution localization feasible. We also derive the Cram\'er-Rao bound for this model, which can be used to guide the placement of collections of receivers so as to optimize localization accuracy. The method improves a balance of accuracy and robustness to environmental model mismatch, relative to existing localization methods that are useful in similar settings. The method is validated with simulations and water tank experiments. 
\end{abstract}


\vspace{-0.3cm}
\section{Introduction}\label{sec:intro}
\vspace{-0.1cm}
Underwater localization of acoustic sources is an important and challenging problem, and arises in a wide range of applications \cite{bahr2009cooperative,corke2007experiments,waterston2019ocean}. As such, it has been extensively addressed in the literature, where early work dates back to at least the mid-$1970$s \cite{bucker1976use}. Fruitful combinations of advanced signal processing methods and detailed underwater acoustic propagation models have led to a variety of methods for different regimes (shallow/deep water, short/long distances, etc.) \cite{tan2011survey,chandrasekhar2006localization}.

While an abundance of methods have been developed and proposed over the years, only a portion of these survive the ruthless test of practicality. Indeed, from a practical point of view, a good \emph{applicable} method is one that, on the one hand exploits as much prior knowledge as possible, but on the other hand does not go too far by assuming access to unavailable information/resources. In the context of passive underwater acoustic localization, our goal in this work is to provide a robust algorithm, while judiciously balancing this trade-off.

In particular, we consider scenarios where the area of interest is characterized by shallow waters (say, up to $\sim100$ m depth \cite{etter1995underwater}) and relatively short distances (say, up to $\sim1$ km). In this regime, under a few additional realistic assumptions (stated explicitly in the sequel), the acoustic signal propagation can be approximated by ray trace modeling \cite{etter1995underwater,jensen2011computational}. This allows us to \emph{exploit} the multipath channel effect, rather than \emph{mitigate} it. In other words, we explicitly incorporate prior knowledge on the structure of the environment, which either allows us to successfully localize using fewer resources (e.g., sensors or measurements), or to improve performance while using the same resources. Moreover, we are capable of localizing a source in the complete absence of line-of-sight (LOS) signal components, based on non-LOS (NLOS) signal reflections. Naturally, these notions have already been considered in some settings, as reviewed in what follows.

\vspace{-0.45cm}
\subsection{Related Work: Underwater Acoustic Localization}\label{subsec:underwaterlocalization}
\vspace{-0.1cm}
For short-range localization in shallow-water environments, straight-ray tracing is a widely-accepted approximation for acoustic signal propagation \cite{etter1995underwater,jensen2011computational}. In such environments, the speed of sound is (at least approximately) constant\footnote{Nearly constant sound speed may be found, e.g., in very shallow waters, or shallow waters that are well-mixed \cite{chitre2007high}.} and known in the relevant volume of interest. Therefore, propagation delay, namely time of arrival (TOA) or time-difference of arrival (TDOA) (e.g., \cite{cheng2008silent,kouzoundjian2017tdoa}), is usually employed as a basis for different localization methods \cite{tuna2017survey}. However, since the underwater acoustic environment typically induces a rich multipath channel \cite{stojanovic1999underwater}, the measured signals contain both LOS and NLOS components. Under such circumstances, the performance of TOA/TDOA-based methods usually deteriorates, possibly up to unacceptable error levels. 

Since the complete multipath channel is (generally) unknown, a possible remedy is to first identify and separate the LOS components. Diamant, \etal\ propose in \cite{diamant2012and} a method for classifying the signal components as LOS and NLOS, and for subsequent range estimation based on the classified LOS components. While this approach can certainly work, it does not attempt to exploit the NLOS reflections, which contain valuable information on the unknown source location. Emokpae and Younis propose in \cite{emokpae2011surface} a surface-reflection-based method in an active setting, where only the surface reflections are exploited. In \cite{emokpae2014ureal}, Emokpae, \etal\ present an extended, enhanced version of this notion, where a scheme that employs both the LOS and surface-reflected NLOS components is developed to locate a lost (drifted away) node of an underwater sensor network. To use this method, all nodes in the network are required to have a sensor array, with more than one sensor, and the waveform emitted from the lost node (i.e., source) is assumed to be known, which is not \delra{necessarily}\addra{always} possible and less common in passive settings. Assuming that perfect knowledge of the physical model is available, which translates into an equivalent impulse response, matched field processing (MFP) \cite{baggeroer1993overview} is a well-known technique that makes full use of the environmental structure for enhanced localization. However, as mentioned in \cite{collins1991focalization}, in realistic applications model mismatch is a serious problem for MFP, on top of its heavy computational workload. Recent increasing efforts towards reducing system cost \cite{iscar2017low,gerondeau2020low} and computational complexity \cite{mantzel2012compressive,gemba2017adaptive}, while exploiting environmental structure \cite{dubrovinskaya2019bathymetry}, motivate our current work.
\vspace{-0.35cm}
\subsection{Semi-Blind Localization: Motivation and Contributions}\label{subsec:motivationandcontributions}
\vspace{-0.1cm}
We propose a semi-blind localization (SBL) method that uses a spatially diverse network of receivers. Each receiver is required to have a single sensor (rather than a sensor array\addra{, as in \cite{emokpae2014ureal}}), which reduces hardware requirements, and hence the overall cost of the system.\footnote{In an application such as the ocean-of-things \cite{waterston2019ocean}, an optimal subset of sensors could be chosen from a larger set of sensors \cite{saucan2020information}.} The information lost by restricting the number of sensors is mitigated by leveraging available partial prior knowledge on the structure of the environment, namely the depths of the sensors and the ocean bottom. Our SBL method, developed in a nonBayesian framework, jointly estimates the associated parameters of the implied impulse response with the unknown source position, and thus can be viewed as a form of focalization \cite{collins1991focalization}. However, it is more \addra{naturally} related to the direct position determination approach \cite{weiss2004direct}, originally proposed for narrowband radio frequency signals. We show that with some carefully \addra{chosen} adaptations, a similar, though generalized approach leads to our SBL method,\footnote{In contrast to the previous claim in \cite{weiss2004direct}, that this approach ``is suitable only for RF signals and not for underwater emitter location".} which provides a good balance between accuracy and robustness to some physical model mismatch. We demonstrate this via simulation experiments by comparing to MFP and to the TDOA method referred to as \addra{``}generalized cross-correlation with phase transform\addra{''} (GCC-PHAT) \cite{knapp1976generalized,brandstein1997robust,grondin2018study}, which is well-known due to its resilience to multipath. 

We note in passing that if additional knowledge of the environment is available, one may consider taking a Bayesian approach, and incorporate the available knowledge by introducing an appropriate prior distribution on (all or some of) the unknowns. In this work, we take a nonBayesian approach.

Our main contributions are the following:
\begin{itemize}
\item \emph{A novel SBL method for underwater acoustic sources}: We adopt the widely-accepted straight-ray tracing approach for shallow-water to define a three-ray model, which explicitly takes into account the NLOS surface and bottom signal reflections. Consequently, on top of enhanced accuracy due to this multipath model, our method is capable of localization in the absence of LOS, due to a potential occluder, such as a vessel or pier pilings.\addra{
\item \emph{Computationally efficient direct localization}: Contrary to indirect (e.g., TDOA-based), standard localization methods (e.g., \cite{wang2019robust,zou2020tdoa,xiong2021tdoa}), we take a different approach, in which our algorithm is applied directly to the observed signals. Consequently, the notion of TDOA is redundant in our framework. Specifically, we assume that the source's waveform is unknown, and in particular, we do not assume it is a pulse-type signal. We provide a computationally efficient algorithm to the resulting nonlinear optimization problem (see Section \ref{sec:proposedsolution}, Proposition \ref{efficientcomputationmaxeigenval}), and demonstrate that the algorithm works well for pulse- or non-pulse-type signals in Section \ref{sec:simulresults}.
}
\item \emph{Lower bound on asymptotic performance}: We develop the Cram\'er-Rao lower bound (CRLB) on the mean-squared error (MSE) of any unbiased localization method for a special case of our signal model, in which our proposed solution coincides with the maximum likelihood estimate (MLE) of the source position. We demonstrate the validity of this bound with respect to ocean ambient noise, using previously collected ocean acoustic recordings \cite{hodgkiss2012kauai}. 
\item \emph{Applicability proof of concept}: We provide a proof of concept, demonstrated on acoustic measurements collected in a well-controlled, small-scale water tank, which provides an acoustically frequency-scaled model for the shallow-water environment. 
\end{itemize}

\begin{figure}[t]
	\includegraphics[width=0.48\textwidth]{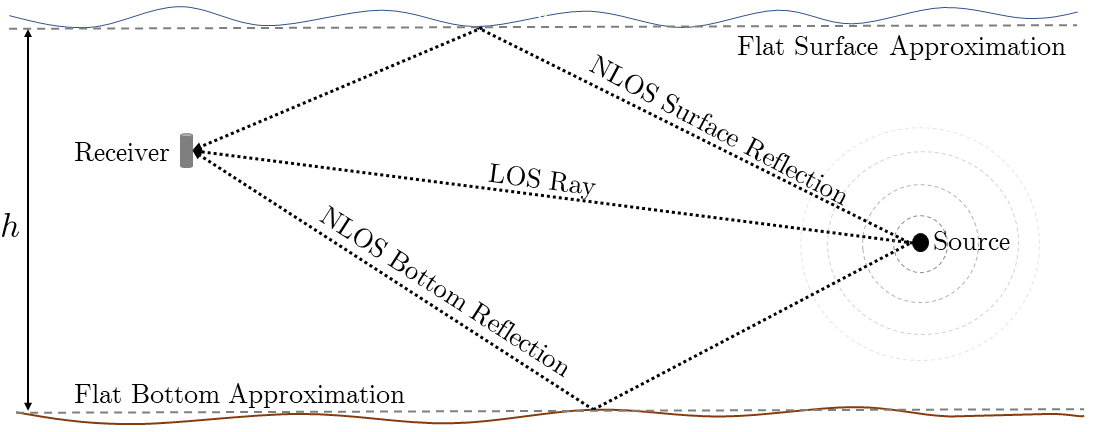}
	\centering
	\caption{A $2$-dimensional illustration of the three-ray model. When the surface and bottom are approximately flat in the operational area\addra{ \cite{etter2018underwater,hovem2011ray}}, this model enables NLOS-based localization in the potential absence of the LOS signal component, e.g., due to an occluder.}
	\label{fig:three_ray_model_illustration}\vspace{-0.4cm}
\end{figure}
The rest of the paper is organized as follows. The remainder of this section is devoted to an outline of our notation. In Section~\ref{sec:problemformulation} we formulate the problem for the three-ray signal model depicted in Fig.~\ref{fig:three_ray_model_illustration}, and the MFP solution of this model is presented in Section~\ref{sec:mfpsolution}. The main results, including our proposed SBL method, are presented in Section~\ref{sec:proposedsolution}. In Section~\ref{sec:cramerraolowerboundGaussian} we derive the respective CRLB, and present empirical simulation results that corroborate our analytical derivation in Section \ref{sec:simulresults}. Concluding remarks are provided in Section~\ref{sec:conclusion}.

\vspace{-0.35cm}
\subsection{Notation}\label{subsec:notations}
\vspace{-0.1cm}
We use $x$, $\ux$, and $\X$ for a scalar, column vector and matrix, respectively. The superscripts $(\cdot)^{\tps}$, $(\cdot)^*$, $(\cdot)^{\her}$, and $(\cdot)^{-1}$ denote the transposition, complex conjugation, conjugate transposition, and inverse operators, respectively. We use $\I_{K}$ to denote the $K\times K$ identity matrix, and $\mathbf{O}$ for the all-zeros matrix. The pinning vector $\ue_k$ denotes the $k$-th column of the identity matrix, with context-dependent dimension. Further, $\delta_{k\ell}\triangleq\ue_k^{\tps}\ue_{\ell}$ denotes the Kronecker delta of $k$ and $\ell$. $\Eset[\cdot]$ denotes expectation, $\lambda_{\max}\left(\cdot\right)$ denotes the largest eigenvalue of its (square) matrix argument, and the $\Diag(\cdot)$ operator forms an $M\times M$ diagonal matrix from its $M$-dimensional vector argument. The Kronecker product is denoted by $\otimes$. We use $\jmath$ (a dotless $j$) to denote $\sqrt{-1}$; $\Re\{\cdot\}$ and $\Im\{\cdot\}$ denote the real and imaginary parts (respectively) of their complex-valued arguments. The $\ell^2$ norm is denoted by $\norm{\cdot}_{2}$, and $\rank(\cdot)$ denotes the rank of its matrix argument. The symbols $\Rset$ and $\Cset$ denote the real line and complex plane, respectively. We use $\mybar{\ux}$ to denote the (normalized) discrete Fourier transform (DFT) of $\ux$, and $\widehat{\ux}$ to denote an estimate thereof. We use $\mathcal{O}(\cdot)$ to denote the standard big O notation \cite{cormen2009introduction}.

\vspace{-0.35cm}
\section{Problem Formulation}\label{sec:problemformulation}
\vspace{-0.1cm}
Consider $L$ spatially-diverse, time-synchronized receivers at known locations, each consisting of a \emph{single} omni-directional hydrophone.\addra{\footnote{\addra{We focus on the single sensor case for convenience. However, our methodology can in principle be used when the receivers have sensor arrays.}}} Furthermore, consider the presence of an unknown signal in an isotropic homogeneous medium, emitted from a source whose deterministic, unknown position is denoted by the vector of coordinates $\up\in\Rset^{3\times 1}$. We assume that the source is static, and is located sufficiently far from all $L$ receivers to permit a planar wavefront (far-field) approximation in the shallow-water waveguide. \addra{Each receiver records the measured acoustic signal on a fixed observation time interval, which after sampling and baseband conversion amounts to $N$ samples. }\delra{Moreover, w}\addra{W}e further assume that the area of operation can be considered as a shallow-water environment, and that the ocean floor depth in the relevant area of operation\footnote{The smallest rectangular area encompassing the source and receivers.} is approximately constant\addra{ \cite{aubauer2000one}}. We restrict our scope to (approximately) isovelocity environments and high frequency signals, in which the straight-ray model approximately holds. Since we focus on short ranges in shallow-water environments, we neglect nonlinear propagation effects in the waveguide.

\begin{figure}[t]
  \centering
  \includegraphics[width=0.95\linewidth]{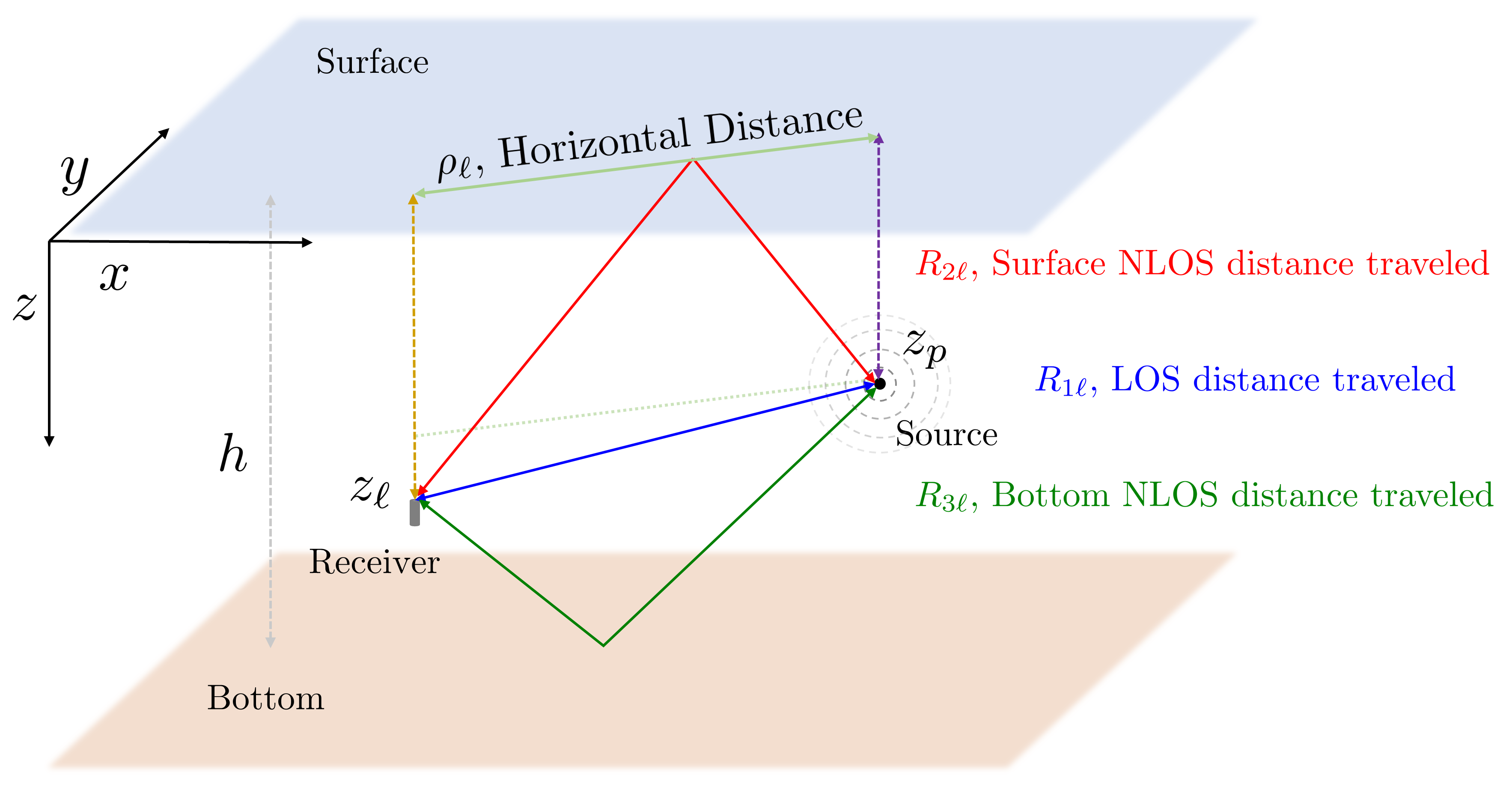}
  \caption{A 3-dimensional illustration of the geometry leading to \eqref{LOSdistance}--\eqref{NLOSbottomdistance}.}\vspace{-0.3cm}
  \label{fig:geometry3D}
\end{figure}

Although the underwater acoustic channel generally gives rise to an equivalent rich multipath channel\delra{ response}, a relatively simple, yet useful, approach is the \emph{three-ray model}, illustrated in Fig.~\ref{fig:three_ray_model_illustration}. In this approach, the modeled signal components are:
\begin{enumerate}
	\itemsep0.2em 
	\item The direct-path LOS component;
	\item The surface reflection NLOS component; and
	\item The bottom reflection NLOS component.
\end{enumerate}
Accordingly, the associated distances traveled by these components from the source to the $\ell$-th receiver are given by\addra{ \cite{1282692}}
\begin{align}
R_{1\ell}&\triangleq\|\up_{\ell}-\up\|_2,\text{ \quad\quad\quad\quad\quad\quad\; (LOS)} \label{LOSdistance}\\
R_{2\ell}&\triangleq\sqrt{\rho_{\ell}^2 + (z_p + z_{\ell})^2},\text{ \quad\quad\;\;\;\, (NLOS surface)} \label{NLOSsurfacedistance}\\
R_{3\ell}&\triangleq\sqrt{\rho_{\ell}^2 + (2h -z_p - z_{\ell})^2},\text{ \quad  (NLOS bottom)} \label{NLOSbottomdistance}
\end{align}
where $\up\triangleq[x_p \; y_p \; z_p]^{\tps}$; $\up_{\ell}\triangleq[x_{\ell} \; y_{\ell} \; z_{\ell}]^{\tps}$ is the position of the $\ell$-th receiver; $\rho_{\ell}\triangleq\sqrt{(x_p - x_{\ell})^2 + (y_p - y_{\ell})^2}$\addra{ is the horizontal distance between the source and the $\ell$-th receiver}; and $h$ is the bottom depth in the area of interest.\addra{ An illustration of the geometry associated with \eqref{LOSdistance}--\eqref{NLOSbottomdistance} in our coordinate system is given in Fig.~\ref{fig:geometry3D}.} Therefore, assuming isovelocity, the associated time-delays of these components are
\begin{equation}\label{timedelyaformula}
\tau_{r\ell}(\up)\triangleq\frac{R_{r\ell}}{c},\; r\in\{1,2,3\}, \; \forall \ell\in\{1,\ldots,L\},
\end{equation}
where $c$ denotes the speed of sound, assumed to be known.

This model can be viewed as a third-order approximation (with respect to the \delra{taps}\addra{delayed signal components}) of the equivalent impulse response of an acoustic channel, whose \delra{total }energy is concentrated in the three \delra{taps}\addra{arrivals} corresponding to the LOS component, and the surface and bottom reflections. While some unpredictable factors can give rise to additional \delra{taps}\addra{components} in the induced impulse response, the surface and bottom of the ocean are always present. Therefore, it is reasonable to incorporate these additional signal components into the model. Moreover, this simplified model allows for successful localization in the absence of (even all) LOS components in the received signals, a situation that may occur, e.g., due to the presence of potential occluders. This will be demonstrated via simulations and\delra{ an} experiment\addra{s} with real data in Section \ref{sec:simulresults}.

\addra{\vspace{-0.3cm}\subsection{Baseband Signal Model}\label{subsec:basebandsignalmodel}}
Formally, and assuming the source has been detected in a given frequency band, the sampled, baseband-converted signal from the $\ell$-th receiver is given by
\begin{equation}\label{modelequation}
\begin{gathered}
x_{\ell}[n]=\sum_{r=1}^{3}b_{r\ell}s_{r\ell}[n]+v_{\ell}[n]\triangleq\us_{\ell}^{\tps}[n]\ub_{\ell}+v_{\ell}[n]\in\Cset,\\
\forall n\in\{1,\ldots,N\}, \;\forall \ell\in\{1,\ldots,L\},
\end{gathered}
\end{equation}
where we have defined $\us_{\ell}[n]=[s_{1\ell}[n]\;s_{2\ell}[n]\;s_{3\ell}[n]]^{\tps}\in\Cset^{3\times1}, \ub_{\ell}=[b_{1\ell}\;b_{2\ell}\;b_{3\ell}]^{\tps}\in\Cset^{3\times1}$, and where
\begin{enumerate}
	\item $b_{r\ell}\in\Cset$ denotes the unknown attenuation coefficient from the source to the $\ell$-th sensor associated with the $r$-th component (LOS or surface/bottom NLOS reflection);
	\item $s_{r\ell}[n]\triangleq \left.s\left(t-\tau_{r\ell}(\up)\right)\right\vert_{t=nT_s}\in\Cset$ denotes the sampled $r$-th component of the unknown signal waveform at the $\ell$-th sensor, where $s\left(t-\tau_{r\ell}(\up)\right)$ is the analog, continuous-time waveform delayed by $\tau_{r\ell}(\up)$, and $T_s$ is the (known) sampling period; and
	\item $v_{\ell}[n]\in\Cset$ denotes the additive noise at the $\ell$-th receiver, representing the overall contributions of internal receiver noise and ambient noise, modeled as a zero-mean random process with an unknown variance $\sigma_{v_{\ell}}^2$.
\end{enumerate}

\subsection{Equivalent Formulation in the Frequency Domain}\label{subsec:freq_formulation}
Applying the normalized DFT to \eqref{modelequation} yields the equivalent frequency-domain representation for all $\ell\in\{1,\ldots,L\}$,
\begin{equation}\label{modelequationfreq}
\begin{aligned}
\hspace{-0.1cm}\xbar_{\ell}[k]&=\sum_{r=1}^{3}b_{r\ell}\sbar[k]e^{-\jmath\omega_k\tau_{r\ell}({\text{\boldmath $p$}})}+\vbar_{\ell}[k]\\
&\triangleq\sbar[k]\cdot\underbrace{\ud_{\ell}^{\her}[k]\ub_{\ell}}_{\triangleq\bar{h}_{\ell}[k]}+\vbar_{\ell}[k]=\sbar[k]\cdot\bar{h}_{\ell}[k]+\vbar_{\ell}[k]\in\Cset,
\end{aligned}
\end{equation}
where we have defined 
\begin{equation*}\label{dft_freq_vec}
\begin{aligned}
\ud_{\ell}[k]&\triangleq[e^{-\jmath\omega_k\tau_{1\ell}({\text{\boldmath $p$}})} \; e^{-\jmath\omega_k\tau_{2\ell}({\text{\boldmath $p$}})} \; e^{-\jmath\omega_k\tau_{3\ell}({\text{\boldmath $p$}})}]^{\her}\in\Cset^{3\times 1},\\
\omega_k&\triangleq\frac{2\pi(k-1)}{NT_s}\in\Rset_+, \;\forall k\in\{1,\ldots,N\}.
\end{aligned}
\end{equation*}

For shorthand, we further define
\begin{equation}\label{initialdefinitionsofsignals}
\begin{aligned}
\bux_{\ell}&\triangleq \left[\mybar{x}_{\ell}[1] \cdots \mybar{x}_{\ell}[N]\right]^{\tps}\in\Cset^{N\times1}, \addra{\bX_{\ell}}\triangleq\Diag(\bux_{\ell}),\\
\bus&\triangleq \left[\sbar[1] \cdots \sbar[N]\right]^{\tps}\in\Cset^{N\times1}, \quad\;\bS\triangleq\Diag(\bus),\\
\buv_{\ell}&\triangleq \left[\mybar{v}_{\ell}[1] \cdots \mybar{v}_{\ell}[N]\right]^{\tps}\in\Cset^{N\times1}, \\
\D_{\ell}&\triangleq\left[\ud_{\ell}[1] \cdots \ud_{\ell}[N]\right]^{\tps}\in\Cset^{N\times 3}, \H_{\ell}\triangleq\Diag\left(\D_{\ell}\ub_{\ell}\right).
\end{aligned}
\end{equation}
Note that $\D_{\ell}$ and $\H_{\ell}$ are \addra{nonlinear }functions of the unknown \addra{emitter}\delra{transmitter} position $\up$,\addra{ as suggested by the definition of $\ud_{\ell}[k]$ above and \eqref{LOSdistance}--\eqref{timedelyaformula},} though we omit this for brevity. With this notation, we may now write \eqref{modelequationfreq} compactly as
\begin{equation}\label{signalmodelfreqcompact}
\bux_{\ell}=\H_{\ell}\bus+\buv_{\ell}\in\Cset^{N\times 1}, \;\forall\ell\in\{1,\ldots,L\}.
\end{equation}

Thus, the localization problem can be formulated as follows:\vspace{0.1cm}
\noindent\textbf{Problem:} {\myfontb\emph{Given the measurements $\left\{\bux_{\ell}\in\Cset^{N\times 1}\right\}_{\ell=1}^{L}$ of the signal model \eqref{signalmodelfreqcompact}, localize the source, namely estimate $\up$.}}\vspace{0.1cm}
We emphasize that although we are interested solely in $\up$, the channel parameters $\{\ub_{\ell}\}$ and the DFT coefficients $\bus$ of the emitted waveform are unknown as well.

\section{The Matched Field Processing Solution}\label{sec:mfpsolution}
The key assumption of MFP approaches is that, for a given hypothesized emitter location $\up$, the channel response $\H_{\ell}$ is \emph{fully} predictable.\footnote{Otherwise, infeasible high-dimensional optimization is required.} For the model \eqref{modelequation}, the attenuation coefficients $\{\ub_{\ell}\}$ can be assumed to be given by\footnote{Ignoring the effects of volume absorption in water, which are minimal.}
\begin{align}
b_{1\ell}&=\frac{1}{R_{1\ell}},\text{ \quad (LOS attenuation)} \label{LOSatten} \\
b_{2\ell}&=\frac{-1}{R_{2\ell}},\text{ \quad (NLOS surface reflection attenuation)} \label{surfaceamplitude}\\
b_{3\ell}&=\frac{\kappa_b}{R_{3\ell}},\text{ \quad (NLOS bottom reflection attenuation)} \label{bottomamplitude}
\end{align}
for all $\ell\in\{1,\ldots,L\}$, where $\kappa_b$ is the bottom reflection coefficient, which (presumably) can be determined based on prior physical knowledge (e.g., assuming the bottom is sand, silt, clay, rock, etc.) and the angle of incidence, and is assumed to be known within the MFP framework for a given hypothesized emitter location $\up$. For \eqref{surfaceamplitude}, we assumed a perfectly reflecting ocean surface \cite{jensen2011computational,emokpae2014surface}, which approximately holds for calm shallow waters. Based on this knowledge, the channel responses $\{\H_{\ell}\}$ can be readily computed.

The MFP solution for the three-ray model, denoted for convenience as MFP3, is then given by
\begin{equation}\label{MFPoptimmization}
\hup_{\MFP} \triangleq \underset{\text{\boldmath$p$}\in\Rset^{3\times1}}{\argmin} \; 
\min_{\text{\boldmath$\bar{s}$}\in\Cset^{N\times 1}}\widetilde{C}_{\MFP}(\up,\bus),
\end{equation}
where
\begin{equation}\label{MFPobjectivefunction}
\widetilde{C}_{\MFP}(\up,\bus)\triangleq\sum_{\ell=1}^{L}\norm{\bux_{\ell}-\H_{\ell}(\up)\bus}_2^2,
\end{equation}
and here we write $\H_{\ell}(\up)$ (rather than $\H_{\ell}$) to emphasize the dependence on $\up$. The simplified MFP3 solution is given by
\begin{equation}\label{MFPsimpleform}
\hup_{\MFP}=\underset{\text{\boldmath$p$}\in\Rset^{3\times1}}{\argmax} \;  \sum_{k=1}^{N}\frac{\left|\bux[k]^{\her}\mybar{\uh}_k(\up)\right|^2}{\|\mybar{\uh}_k(\up)\|^2},
\end{equation}
where we have defined, for every $k$-th DFT component,
\begin{align*}
\bux[k]&\triangleq\left[\bar{x}_1[k]\,\cdots\,\bar{x}_L[k]\right]^{\tps}\in\Cset^{L\times 1},\\
\mybar{\uh}_k(\up)&\triangleq\left[\bar{h}_{1}[k]\;\cdots\;\bar{h}_{L}[k]\right]^{\tps}\in\Cset^{L\times 1},
\end{align*}
using $\mybar{\uh}_k(\up)$ (rather than $\mybar{\uh}_k$) to emphasize the dependence on $\up$. For completeness of the exposition, the derivation of the simplified form \eqref{MFPsimpleform} is given in the supplementary materials. 

In \eqref{MFPsimpleform}, the channel impulse response $\mybar{\uh}_k(\up)$ is considered to be fully known for any given hypothesized position $\up$ (via \eqref{LOSdistance}--\eqref{timedelyaformula} and \eqref{LOSatten}--\eqref{bottomamplitude}). In other words, assuming perfect knowledge of $\ub_{1},\ldots,\ub_{L}$ means that any relevant physical parameter, such as the ocean bottom sediment coefficient $\kappa_b$ in \eqref{bottomamplitude}, is assumed to be perfectly known as well. We relax this (somewhat unrealistic) assumption in our semi-blind localization approach described next.

\vspace{-0.1cm}
\section{The Proposed Semi-Blind Localization Method}\label{sec:proposedsolution}
\delra{In}\addra{As} our semi-blind framework, \addra{we only assume that the bottom depth $h$ is known, but }we do not assume \delra{to}\addra{that we} have any prior knowledge of the channel attenuation coefficients. Thus, since the waveform emitted from the source is also unknown, we may assume without loss of generality (w.l.o.g.) that $\|\bus\|_2=1$, viz., $\bus\in\mathcal{S}_N\triangleq\{\uz\in\Cset^{N\times1}\colon\|\uz\|_2=1\}$, where $\mathcal{S}_N$ is the $N$-dimensional unit sphere.\addra{ This assumption, which is common in similar (semi-)blind formulations (e.g., \cite{9293153}), is justified due to the inherent \emph{scaling} ambiguity in \eqref{signalmodelfreqcompact},
\begin{align*}
    \alpha\in\Cset:\H_{\ell}\bus&=\Diag\left(\D_{\ell}\ub_{\ell}\right)\bus=\Diag\Big(\D_{\ell}\big(\underbrace{\tfrac{1}{\alpha}\ub_{\ell}}_{\triangleq\,\widetilde{\ub}_{\ell}}\big)\Big)(\underbrace{\alpha\bus}_{\triangleq\,\widetilde{\bus}})\\
    &=\underbrace{\Diag\left(\D_{\ell}\widetilde{\ub}_{\ell}\right)}_{\triangleq\,\widetilde{\H}_{\ell}}\widetilde{\bus}=\widetilde{\H}_{\ell}\widetilde{\bus}, \quad \forall\ell\in\{1,\ldots,L\},
\end{align*}
which, granted, is immaterial to our localization problem.}

Our proposed SBL solution can be viewed as the MLE of $\up$, obtained by joint estimation of \emph{all} the unknown deterministic model parameters, under the assumption that the noise processes $\{\buv_{\ell}\}_{\ell=1}^{L}$ from all different sensors are temporally white complex normal (CN) processes, mutually statistically independent, with equal variances. In this case, the MLE of $\up$ is the solution to the nonlinear least squares problem
\begin{equation}\label{MLEoptimmization}
\hup_{\SBL} \triangleq \underset{\text{\boldmath$p$}\in\Rset^{3\times1}}{\argmin} \; 
\underbrace{\min_{\substack{\text{\boldmath$\bar{s}$}\in\mathcal{S}_N \\ {\text{\boldmath$B$}}\in\Cset^{3\times L}}} \widetilde{C}_{\SBL}(\up,\bus,\B)}_{C_{\SBL}(\text{\boldmath$p$})} \triangleq
\underset{\text{\boldmath$p$}\in\Rset^{3\times1}}{\argmin} \; C_{\SBL}(\up),\\
\end{equation}
where the objective function $\widetilde{C}_{\SBL}(\up,\bus,\B)$ is defined as
\begin{equation}\label{NLSobjectivefunction}
\widetilde{C}_{\SBL}(\up,\bus,\B)\triangleq \sum_{\ell=1}^{L}\norm{\bux_{\ell}-\Diag\left(\D_{\ell}(\up)\ub_{\ell}\right)\bus}_2^2,
\end{equation}
and here we write $\D_{\ell}(\up)$ (rather than $\D_{\ell}$) to emphasize the dependence on $\up$. In contrast to the MFP3 solution \eqref{MFPoptimmization}, our proposed solution \eqref{MLEoptimmization} is due to joint estimation of all the unknown model parameters $\up, \bus$ and $\B\triangleq\left[\ub_{1} \ldots \ub_{L}\right]\in\Cset^{3\times L}$, including the channel coefficients $\B$. Thus, in our proposed approach, we do \emph{not} assume that the channel response is fully known for a given hypothesized position $\up$ of the source. 

Intuitively, this approach should lead to a more robust solution than MFP3 with respect to deviations from the channel knowledge \eqref{LOSatten}--\eqref{bottomamplitude}, at the cost of extra computational effort. Fortunately, as we show in Section \ref{sec:proposedsolution}, by exploiting the low-dimensional structure of the data, the additional computational cost is negligible. Moreover, although \eqref{MLEoptimmization} defines a nonlinear high-dimensional optimization problem with $3L$ additional unknowns relative to MFP3, it boils down to a $3$-dimensional optimization problem, similar to \eqref{MFPsimpleform}.

Our main result is the following localization algorithm:
\tcbset{colframe=gray!90!blue,size=small,width=0.49\textwidth,halign=flush center,arc=2mm,outer arc=1mm}
\begin{tcolorbox}[upperbox=visible,colback=white,halign=left]
	\textbf{\underline{The SBL Estimator:}}\\
	\textbf{Input}: $\left\{\bux_{\ell}\right\}_{\ell=1}^{L}, c, h$, $3$D grid of the volume of interest. \\ \textbf{Output}: The SBL estimate, $\hup_{\SBL}$.
	\begin{enumerate}[1.]
		\item For every candidate $\up$ on the grid:
		\begin{enumerate}
			\item[1.1.] Compute the matrices $\{\D_{\ell}^{\tps}\D_{\ell}^* \}_{\ell=1}^L$;
			\item[1.2.] Compute the Cholesky decompositions
			\begin{equation}\label{Cholesky}
                \D_{\ell}^{\tps}\D^*_{\ell}\triangleq\mGamma_{\ell}^{\her}\mGamma_{\ell}\in\Cset^{3\times 3}, \; \forall \ell\in\{1,\ldots,L\},
            \end{equation}
			and obtain the matrices $\{\mGamma_{\ell} \}_{\ell=1}^L$;
			\item[1.3.] Compute the matrix
			\begin{equation}\label{Umatrices}
                \U(\up)\triangleq\left[\bX_{1}\D^*_{1}\mGamma_{1}^{-1} \cdots \; \bX_{L}\D^*_{L}\mGamma_{L}^{-1}\right]\in\Cset^{N\times 3L},
            \end{equation}
			and construct the matrix
			\begin{equation}\label{def_of_Q_tilde}
                \widetilde{\Q}(\up)\triangleq \U(\up)^{\her}\U(\up)\in\Cset^{3L\times3L}.
            \end{equation}
			\item[1.4.] Compute $\lambda_{\max}\left(\widetilde{\Q}(\up)\right)$;
		\end{enumerate}
	\item Find $\hup_{\SBL,\text{grid}}$, the maximizer point on the grid.
	\item Return $\hup_{\SBL}$, the solution of a nonlinear optimization solver (e.g., trust-region \cite{conn2000trust}) initialized by $\hup_{\SBL,\text{grid}}$.
	\end{enumerate}
\end{tcolorbox}

We now provide the \delra{theoretical results}\addra{analysis}, based on which the algorithm above is derived. For convenience, we define
\begin{equation}\label{approxflatsignal}
\P_{\sbar} \triangleq \Diag(|\bus|^2) \triangleq P_s\cdot\left(\I_N + \meps\right)\in\Rset_+^{N\times N},
\end{equation}
where $P_s\in\Rset_+$ is the average signal power (with $|\cdot|^2$ operating elementwise), and $\meps$ is a diagonal matrix with ``small" elements, such that $\varepsilon_{\max}\triangleq|\lambda_{\max}\left(\meps\right)|< 1$.

\begin{prop}[SBL for Spectrally Flat Waveforms]\label{maxeigenvalueflatspectrum}
\delra{\emph{[\textit{SBL for Spectrally Flat Waveforms}]}}  Consider the case where $\meps=\mathbf{O}$, and define the data-dependent matrix,
\begin{equation}\label{targetmatrixdef}
\Q(\up)\triangleq\sum_{\ell=1}^{L}\bX_{\ell}\D^*_{\ell}\left(\D_{\ell}^{\tps}\D^*_{\ell}\right)^{-1}\left(\bX_{\ell}\D^*_{\ell}\right)^{\her}\in\Cset^{N\times N},
\end{equation}
for any hypothesized source position $\up$. Then,
\begin{equation}\label{SBLestimate}
\hup_{\emph{\SBL}}=\underset{\text{\boldmath$p$}\in\Rset^{3\times1}}{\argmax} \; \lambda_{\max}\left(\Q(\up)\right).
\end{equation}
\end{prop}
Proposition \ref{maxeigenvalueflatspectrum}, whose proof is given in Appendix \ref{proofofprop1}, tells us that, for spectrally flat waveforms $\bus$, the source's position estimator can be computed based only on $\lambda_{\max}\left(\Q(\up)\right)$. Moreover, although our model has more unknowns, \eqref{SBLestimate} is obtained by (only) a $3$-dimensional optimization.

The next proposition, whose proof appears in Appendix \ref{proofofprop2}, states that the simplified form $\lambda_{\max}\left(\Q(\up)\right)$ of the objective function can be a good approximation to \eqref{MLEoptimmization} when $\meps\neq\mathbf{O}$, namely for waveforms that are not spectrally flat. In turn, this implies that \eqref{SBLestimate} can be used to localize a source emitting a general waveform.

\begin{prop}[SBL for General Waveforms]\label{maxeigenvaluenonflatspectrum}
\delra{\emph{[\textit{SBL for General Waveforms}]}}  Consider the case where $\meps$ is not necessarily equal to $\mathbf{O}$. Then,
\begin{equation*}
\hup_{\emph{\SBL}}=\underset{\text{\boldmath$p$}\in\Rset^{3\times1}}{\argmax} \; \lambda_{\max}\left(\Q(\up)\right) + \mathcal{O}(\varepsilon_{\max}).
\end{equation*}
\end{prop}

It follows that whenever $\varepsilon_{\max}\ll 1$, we have
\begin{equation*}
\hup_{\SBL}\approx\underset{\text{\boldmath$p$}\in\Rset^{3\times1}}{\argmax} \; \lambda_{\max}\left(\Q(\up)\right).
\end{equation*}
However, as we demonstrate via simulations and real data in Section \ref{sec:simulresults}, our proposed estimator exhibits good performance even for waveforms that are far from being spectrally flat. Thus, Proposition \ref{maxeigenvaluenonflatspectrum} implies that only  $\lambda_{\max}\left(\Q(\up)\right)$ is required for approximately optimal localization. In particular, it suffices to use, e.g., the power method, rather than computing the complete eigenvalue decomposition of $\Q(\up)$. However, the computational complexity can be reduced even more, as implied by the following proposition, whose proof is given in Appendix \ref{proofofprop3}.

\begin{prop}[Efficient Computation of the SBL Objective Function]\label{efficientcomputationmaxeigenval}\delra{
\emph{[\textit{Efficient Computation of the SBL Objective Function}]}}  Let $\Q(\up)\in\Cset^{N\times N}$ be defined as in \eqref{targetmatrixdef}. Then,
\begin{equation}\label{maxeigenval}
\lambda_{\max}\left(\Q(\up)\right)=\lambda_{\max}\left(\widetilde{\Q}(\up)\right),
\end{equation}
and the complexity of computing \eqref{maxeigenval} is (only) $\mathcal{O}(NL^2)$.
\end{prop}
We note that a na\"ive application of, e.g., the power method to $\Q(\up)$ would cost $\mathcal{O}(N^2)$. This is already prohibitively expensive for reasonable sample sizes on the order of $N\sim10^3$.

We emphasize that our proposed estimator implicitly optimizes over an additional $2\cdot3L$ unknown parameters---the channel attenuation coefficients $\ub_{1},\ldots,\ub_{L}$---relative to the MFP estimator of this model, while retaining the same order of computational complexity in terms of $N$ (sample size). As an intermediate summary, a comparison of several attributes of the proposed SBL with MFP3 is given in Table \ref{table:comparisontoMFP}.

\begin{table}[t]
	\centering
	\noindent\makebox[0.515\textwidth]{
	\begin{tabular}{c|c|c|ll}
		\cline{2-3}	\rule{0pt}{0.5cm}
		& MFP3 & SBL &  &  \\[0.25cm] \cline{1-3}
		\multicolumn{1}{|c|}{Unknowns}    &  $\up,\bus$   &  \rule{0pt}{0.5cm}$\up,\bus, \ub_1,\ldots,\ub_{L}$    &  &  \\[0.25cm] \cline{1-3}
		\multicolumn{1}{|c|}{\begin{tabular}[c]{@{}c@{}}Objective \\ function\end{tabular}}                    &  \rule{0pt}{0.5cm}$\sum_{k=1}^{N}\frac{\left|\bux[k]^{\her}\uh_k(\up)\right|^2}{\|\uh_k(\up)\|^2}$   &   $\lambda_{\max}\left(\widetilde{\Q}(\up)\right)$   &  &  \\[0.25cm] \cline{1-3}
		\multicolumn{1}{|c|}{Complexity}                   &  \rule{0pt}{0.5cm}$\mathcal{O}(NL)$   &   $\mathcal{O}(NL^2)$   &  &  \\[0.25cm] \cline{1-3}
		\multicolumn{1}{|c|}{\begin{tabular}[c]{@{}c@{}}Required \\ physical parameters\end{tabular}} &  \rule{0pt}{0.5cm}$h,c,\kappa_b,\ub_1,\ldots,\ub_{L}$   &  $h,c$    &  &  \\[0.25cm] \cline{1-3}
	\end{tabular}}
	\caption{Comparison of the primary attributes of traditional MFP3 and the proposed estimator, SBL, for the three-ray model.}
	\label{table:comparisontoMFP}\vspace{-0.3cm}
\end{table}

For the actual computation of the estimate $\hup_{\SBL}$, we propose a two-phased approach. The first phase consists of a coarse grid search over the relevant volume of interest. In the second phase, a general purpose nonlinear optimization algorithm (e.g., trust-region methods \cite{conn2000trust}) is applied, where the solution from the first phase is used for initialization.

\vspace{-0.3cm}
\subsection{Interpretation of the SBL Solution}\label{subsec:sblinterpretation}

We now provide a useful interpretation of the closed-form expression \eqref{SBLestimate} of our proposed solution, based on the derivation presented in Appendix \ref{proofofprop1}. We begin by explaining the first step, the estimation of $\ub_{\ell}$. From \eqref{bsolutionLS}, when $\bS$ and $\D_{\ell}$ (defined in \eqref{initialdefinitionsofsignals}) are treated as known, we see that this first step can be regarded as compensation (or, rectification) of the attenuations of each of the three signal components. It is also enlightening to see this from the noiseless case, where
\begin{equation}
\bux_{\ell}=\bS\D_{\ell}\ub_{\ell} \; \Longrightarrow \underbrace{\bS^{-1}\bux_{\ell}}_{\substack{\text{per-frequency} \\ {\text{elementwise division}}}}=\underbrace{\D_{\ell}\ub_{\ell}}_{\substack{\text{per-frequency weighted} \\ {\text{sum of $b_{1\ell},b_{2\ell},b_{3\ell}$}}}}.
\end{equation}
Substituting $\{\widehat{\ub}_\ell\}$ (defined in \eqref{differentformofbest}) into \eqref{NLSobjectivefunction} yields after simplification \eqref{Postivecost}---the ``$\ub_{\ell}$'s-rectified" objective, where the rectification is based on the intermediate estimators $\{\widehat{\ub}_\ell\}$, which still depend on the unknown $\bS$ and $\up$ at this phase. 

Moving forward, we momentarily focus on a single (matrix) element of the sum \eqref{Postivecost}. Rearranging this term, we see that
\begin{align}\label{Qldifferent}
&\bus^{\her}\bX_{\ell}\D^*_{\ell}\left({\D_{\ell}}^{\tps}\P_{\sbar}\D^*_{\ell}\right)^{-1}\left(\bX_{\ell}\D^*_{\ell}\right)^{\her}\bus=\nonumber\\
&\bux_{\ell}^{\tps}\bS^*\D^*_{\ell}\left({\D_{\ell}}^{\tps}\bS^{\tps}\bS^*\D^*_{\ell}\right)^{-1}\D^{\tps}_{\ell}\bS^{\tps}\bux_{\ell}^*.
\end{align}
Again, focusing on the noiseless case to gain intuition, by substituting $\bux_{\ell}=\bS\D_{\ell}\ub_{\ell}$, we have
\begin{align*}
&\bux_{\ell}^{\tps}\bS^*\D^*_{\ell}\left({\D_{\ell}}^{\tps}\bS^{\tps}\bS^*\D^*_{\ell}\right)^{-1}\D^{\tps}_{\ell}\bS^{\tps}\bux_{\ell}^*=\\
&\ub_{\ell}^{\tps}\D_{\ell}^{\tps}\bS^{\tps}\bS^*\D^*_{\ell}\left({\D_{\ell}}^{\tps}\bS^{\tps}\bS^*\D^*_{\ell}\right)^{-1}\D^{\tps}_{\ell}\bS^{\tps}\bS^*\D_{\ell}^*\ub^*_{\ell}=\\
&\ub_{\ell}^{\tps}\D_{\ell}^{\tps}\bS^{\tps}\bS^*\D^*_{\ell}\ub^*_{\ell}=\|\bS\D_{\ell}\ub_{\ell}\|^2=\sum_{k=1}^{N}\left|\sbar[k]\ud_{\ell}^{\her}[k]\ub_{\ell}\right|_2^2.
\end{align*}
Therefore, we interpret the maximization \eqref{Postivecost}---for a single receiver---as choosing the best set of parameters $\{\bus,\up,\ub_{\ell}\}$, in the sense that the total energy of the received signal from the source is maximized, under the hypothesized set of parameters.

Generalizing this intuition for a signal in noise, after substituting $\bux_{\ell}$ into \eqref{Qldifferent}, for a sufficiently large $N$, the signal-noise cross product terms will tend to zero by virtue of the law of large numbers, since the noise DFT coefficients are uncorrelated and zero-mean.

Lastly, we generalize the intuition above from a single receiver to multiple receivers. For this, recall that \eqref{Postivecost} is in fact a \emph{joint} maximization of the total energy of all $L$ received signals from the same source. Therefore, it weights the $L$ signals from different relative locations to the source while taking into account that they all contain shifted versions of \emph{the same} waveform. This is essentially the connecting link, and the advantage in processing the data jointly (rather than individually). This joint weighting is non\delra{-}trivial in the general case. However, when the source is spectrally flat, i.e., $\meps=\mathbf{O}$, the optimal way (in the sense of \eqref{MLEoptimmization}) to weight and combine the data from the receivers is to form the matrix $\Q(\up)$ as in \eqref{targetmatrixdef}, and to compute its maximal eigenvalue \eqref{SBLestimate}. A natural interpretation of the maximal eigenvalue of a semi-positive definite matrix is the energy distributed along the dominant direction (orthogonal to all others) in the space spanned by the columns of this matrix. With this interpretation, the final form of the SBL solution given in \eqref{SBLestimate} is now intuitive.

\vspace{-0.1cm}
\section{The Cram\'er-Rao Lower Bound\addra{ for SBL}}\label{sec:cramerraolowerboundGaussian}
We now analyze the localization accuracy limitations of the proposed solution in terms of the MSE,
\begin{equation}\label{MSEdefinition}
\MSE(\hup,\up)\triangleq\Eset\left[\norm{\hup-\up}_2^2\right].
\end{equation}
Specifically, we derive the CRLB for the special case $\meps=\mathbf{O}$. Unlike the common approach (e.g., as in \cite{baggeroer2017stochastic,baggeroer1988matched}), wherein both the unknown source signal and noise are considered to be random, in our model only the noise is considered random. Thus, for a given waveform, the bound can be used as a tool for designing the deployment of a network of receivers, so as to maximize accuracy in regions of higher importance.

Regardless of the constant spectral level (i.e., $\meps=\mathbf{O}$), in our general framework $\bus\in\mathcal{S}_N$ w.l.o.g., hence in this particular case $P_s=\frac{1}{N}$. Consequently, $\sbar[k]=\frac{1}{\sqrt{N}}e^{\jmath\phi_s[k]}$ for all $k$, and the only waveform-related unknowns are the phases\footnote{Note that although there are $N$ elements in $\uphi_s$, there are only $N-1$ degrees of freedom, since the (complex-valued) channel attenuation coefficients are considered unknown as well. Therefore, we assume w.l.o.g.\ that the first element of $\uphi_s$, considered as a reference phase, is zero.} of the DFT coefficients, denoted collectively by $\uphi_s\triangleq[\phi_s[2] \ldots \phi_s[N]\addra{]}\in\Rset^{(N-1)\times1}$.

To facilitate the following derivation, we introduce a more compact representation of the measured signals. Specifically, let $\bux\triangleq[\bux_1 \ldots \bux_L]^{\tps}\in\Cset^{NL\times 1}$. Thus, \eqref{signalmodelfreqcompact} reads
\begin{equation}\label{compactmodelforCRLB}
\bux=\H\bus+\buv\in\Cset^{NL\times 1},
\end{equation}
where $\H\triangleq\left[\H_1^{\tps} \ldots \H_L^{\tps}\right]^{\tps}\in\Cset^{NL\times N}$ and $\buv\triangleq[\buv_1 \ldots \buv_L]^{\tps}$. Denoting $\usigma_v^2\triangleq[\sigma^2_{v_{1}} \ldots \sigma^2_{v_{L}}]^{\tps}\in\Rset_+^{L\times1}$, it follows that
\begin{equation}\label{CN_vector_model}
\bux\sim \mathcal{CN}\left(\H\bus,\Diag(\usigma^2_v)\otimes\I_{N}\right).
\end{equation}
It is well-known that for the CN signal model $\mathcal{CN}\left(\umu,\R\right)$, the Fisher \delra{I}\addra{i}nformation \delra{M}\addra{m}atrix (FIM) elements are given by\footnote{For the sake of clarity, we specifically use a different notation for the FIM's elements, with slight abuse of notation also in \eqref{FIM_noise}--\eqref{FIM_signal}.} \cite{collier2005fisher}
\begin{equation*}
\begin{gathered}
J[\theta_i,\theta_j]=\Tr\left(\R^{-1}\frac{\partial\R}{\partial\theta_i}\R^{-1}\frac{\partial\R}{\partial\theta_j}\right)+2\Re\left\{\frac{\partial\umu^{\her}}{\partial\theta_i}\R^{-1}\frac{\partial\umu}{\partial\theta_j}\right\},\\
\forall i,j\in\{1,\ldots,K_{\theta}\},
\end{gathered}
\end{equation*}
where we have defined the vector of all the real-valued unknown deterministic parameters
\begin{equation}\label{defoftheta}
\utheta\hspace{-0.05cm}\triangleq\hspace{-0.05cm}\left[\up^{\tps}\,\text{vec}(\uphi_s)^{\tps}\,\text{vec}(\Re\{\B\})^{\tps}\,\text{vec}(\Im\{\B\})^{\tps}\,\usigma_v^2\right]^{\tps}\hspace{-0.15cm}\in\hspace{-0.05cm}\Rset^{K_{\theta}\times1},
\end{equation} 
with $K_{\theta}=3+(N-1)+2\cdot3L+L$, and $\J(\utheta)$ is the FIM.

It is readily seen from \eqref{CN_vector_model}, that in our model the mean vector and covariance matrix are functions of distinct unknown parameters. This immediately implies that
\begin{align}
J[\sigma_{v_{\ell_1}}^2,\sigma_{v_{\ell_2}}^2]&=N\cdot\delta_{\ell_1\ell_2}, \; \forall \ell_1,\ell_2\in\{1,\ldots,L\},\label{FIM_noise}\\
J[\sigma_{v_{\ell}}^2,\theta]&=0, \quad\quad\quad\; \forall \theta\neq\sigma_{v_{\ell}}^2,\label{FIM_cross}
\end{align}
namely the FIM has a block diagonal structure. Furthermore, for the signal-related block, we have
\begin{equation}\label{FIM_signal}
\begin{gathered}
J[\theta_i,\theta_j]=2\Re\left\{\frac{\partial(\H\bus)^{\her}}{\partial\theta_i}\left(\Diag^{-1}(\usigma^2_v)\otimes\I_{N}\right)\frac{\partial\H\bus}{\partial\theta_j}\right\},\\
\forall i,j\in\{1,\ldots,K_{\theta}\}, \; \forall \ell\in\{1,\ldots,L\}.
\end{gathered}
\end{equation}
When $\sigma_{v_{\ell}}^2=\sigma_v^2$ for all $\ell$, it can be observed from \eqref{FIM_signal} that the signal-related FIM block is inversely proportional to the noise variance. Hence, the associated signal-related CRLB block is inversely proportional to the signal-to-noise ratio (SNR).

It only remains to compute the derivatives of $\H\bus$ with respect to the parameters of $\utheta$, excluding $\usigma_v^2$, which is merely technical. We defer the details of these calculations, as well as the final expressions of all the signal-related elements of the FIM to the supplementary materials, along with a Matlab implementation of this bound. Finally, the CRLB is given by\footnote{$\A\succeq\B$ \delra{implies}\addra{is to be interpreted to mean} that $\delra{(}\A-\B\delra{)}$ is semi-positive definite.}
\begin{equation}\label{CRLBonMSEinequality}
\begin{gathered}
\Eset\left[\left(\widehat{\utheta}-\utheta\right)\left(\widehat{\utheta}-\utheta\right)^{\tps}\right]\succeq \J^{-1}(\utheta)\triangleq \CRLB(\utheta)\in\Rset^{K_{\theta}\times K_{\theta}}\\
\Longrightarrow \; \MSE(\hup,\up)\geq \sum_{i=1}^{3}\left[\CRLB(\utheta)\right]_{ii},
\end{gathered}
\end{equation}
for any unbiased estimator $\hutheta$, and the implied $\hup$ (see \eqref{defoftheta}).

\vspace{-0.3cm}
\section{Simulation and Experimental Results}\label{sec:simulresults}
In this section, we consider simulation and physical experiments of source localization for different scenarios in order to corroborate our analytical derivations. First, we begin by the evaluation and visualization of the CRLB for a hybrid signal, wherein the signal-related component, namely $\H\bus$ from \eqref{compactmodelforCRLB}, is synthetic, and the noise-related component, namely $\buv$ from \eqref{compactmodelforCRLB}, is taken from previously collected ambient noise recordings from the Kauai ACOMMS ONR MURI 2011 (KAM11) experiment \cite{hodgkiss2012kauai}. Second, we simulate a different scenario, wherein the receivers are deployed in a linear formation. For this setting, we evaluate the performance with respect to varying SNR, model mismatch (to assess robustness), and missing LOS components due to occluders. In these simulations, we compare our proposed method to the MFP3 solution \eqref{MFPsimpleform} and to GCC-PHAT \cite{brandstein1997robust}, a TDOA-based localization method, which is considered as highly robust to multipath effects. In the third experiment we compare the algorithms on data recorded from a water tank testbed.

%

\subsection{Validation of the CRLB for Ocean Ambient Noise}\label{subsec:CRLBvisualization}
We consider a scenario with $L=4$ receivers, in an area with bottom depth $h=100\;\text{m}$. The locations of the receivers and the source are given in Table \ref{table:positionsforCRLB}. The attenuation coefficients were drawn (once, and then fixed) independently from the circularly-symmetric CN distribution, such that $\Eset\left[|b_{rl}|^2\right]=1$, with variance $0.1^2$. The speed of sound was set to $c=1500\;\text{m}/\text{s}$, and the sample size to $N=30$. We consider the case $\meps=\mathbf{O}$, such that the waveform's DFT coefficients are $\sbar[k]=\frac{1}{\sqrt{N}}e^{\jmath\phi_s[k]}$, and the phases $\{\phi_s[k]\}_{k=2}^N$ were drawn\footnote{Except for the (immaterial) $\phi_s[1]=0$, due to our semi-blind setting.} (once, and then fixed) independently from the uniform distribution $U(0,2\pi)$. \begin{comment}{The magnitudes $|\sbar[k]|=\frac{1}{\sqrt{N}}$ are constant to have \eqref{SBLestimate} as is a strict equality.}\end{comment} In this case, $\hup_{\SBL}$ is the MLE, and the CRLB accurately predicts its asymptotic variance. The received signals were generated according to \eqref{modelequationfreq}, where the noise realization for all four sensors were taken from recordings of ocean ambient noise from the KAM11 experiment \cite{hodgkiss2012kauai}. This way, we obtain a hybrid signal for this simulation, which allows us to test the validity of the bound on real ambient noise, which is potentially not CN and temporally white. Since the CRLB is informative only asymptotically (in the ``small errors" regime), we set the noise variance\footnote{We do so by first normalizing the recorded ambient noise to have unit variance, and then scale it accordingly to have the desired level of SNR.} to $\sigma_{v_{\ell}}^2=\sigma_v^2=0.1$ for all $\ell\in\{1,2,3,4\}$, to have an SNR of $\norm{\bus}_2^2/\sigma_v^2=10$ dB.

\begin{table}[t!]
	\centering
	\noindent\makebox[0.515\textwidth]{
	\begingroup
	\setlength{\tabcolsep}{10pt} 
	\renewcommand{\arraystretch}{1.3} 
	\begin{tabular}{|c|c|c|c|l}
		\cline{1-4}
		& $x$ {[}m{]} & $y$ {[}m{]} & $z$ {[}m{]} &  \\ \cline{1-4}
		Source Position, $\up$ &   $200.7240$      &   $100.1661$         &    $30.6374$        &  \\ \cline{1-4}
		Receiver 1, $\up_1$    &     $150$       &    $-175$        &       $20$     &  \\ \cline{1-4}
		Receiver 2, $\up_2$     &     $75$       &    $-225$        &   $20$         &  \\ \cline{1-4}
		Receiver 3, $\up_3$    &    $-50$        &    $-200$        &       $20$     &  \\ \cline{1-4}
		Receiver 4, $\up_4$    &    $-150$        &   $-150$         &       $20$     &  \\ \cline{1-4}
	\end{tabular}
	\endgroup}
	\caption{Positions of the source and the four receivers for the setting considered in Subsection \ref{subsec:CRLBvisualization}, depicted in Fig.~\ref{fig:CRLB_2D_ellipse} (left), with $T_s=10^{-3}\text{ s}$. Note that the source position is not located on a (discrete) grid point.}
	\label{table:positionsforCRLB}\vspace{-0.1cm}
\end{table}
\begin{figure}[t!]
	\includegraphics[width=0.5\textwidth]{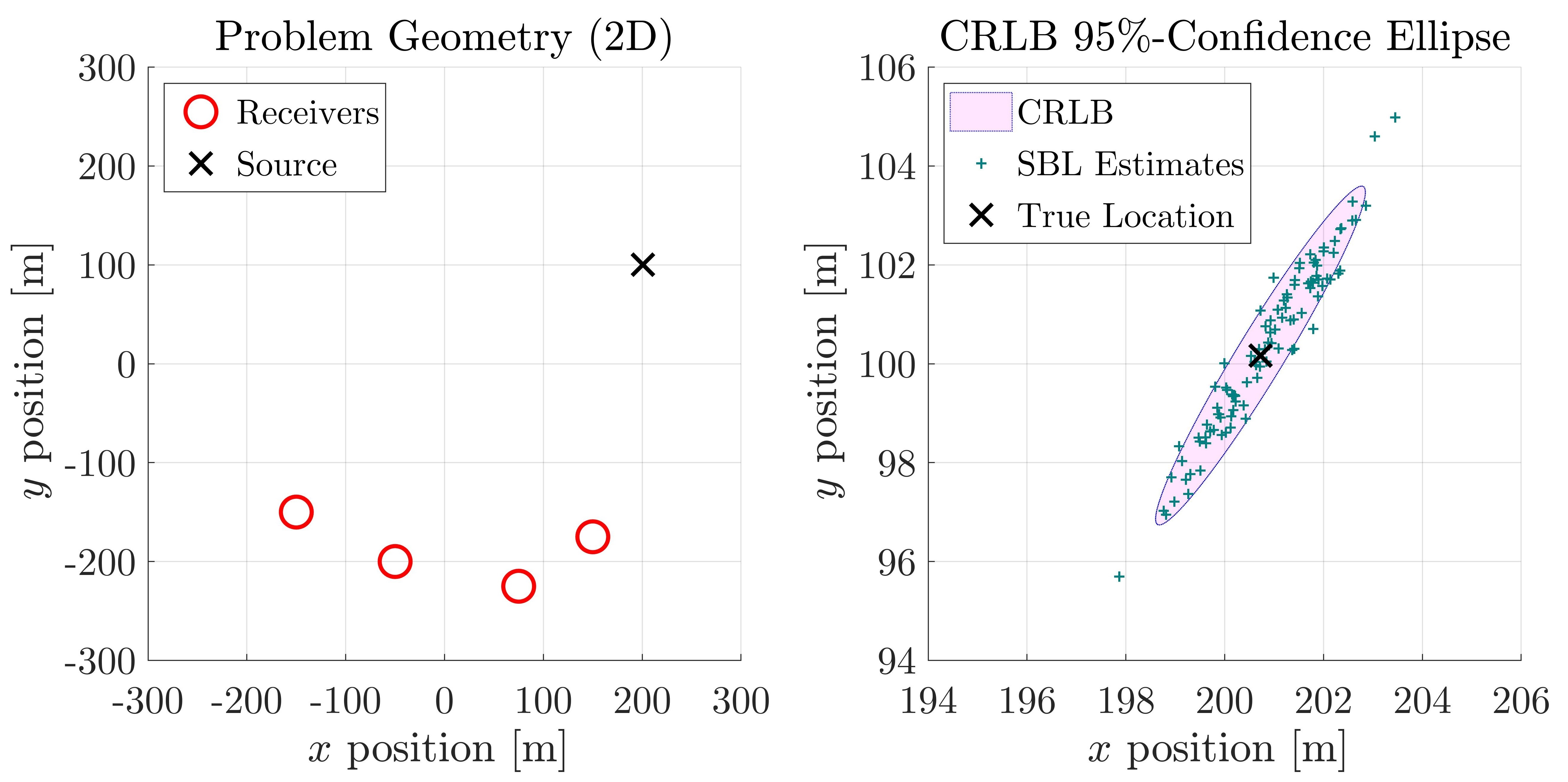}
	\centering
	\caption{Validation of the predicted asymptotic performance using the CRLB for the hybrid signal, containing recordings of ocean ambient noise collected in the KAM11 experiment. Left: The $2$-dimensional setting of the scenario under consideration. Right: The $95\%$-confidence ellipse, as predicted by the CRLB \eqref{CRLBonMSEinequality}, with $100$ superimposed estimates.}
	\label{fig:CRLB_2D_ellipse}\vspace{-0.5cm}
\end{figure}
\begin{figure}[t!]
	\includegraphics[width=0.5\textwidth]{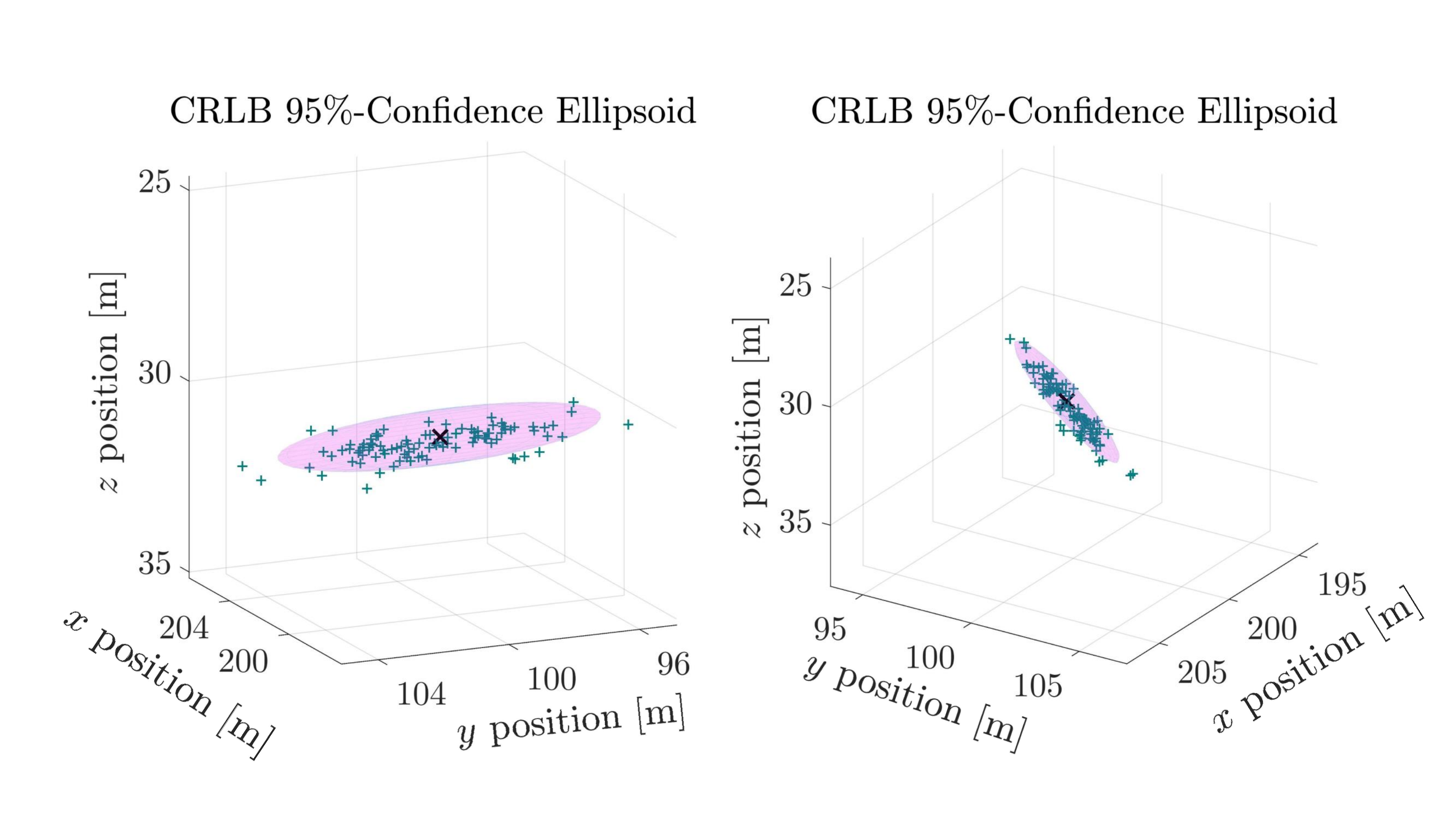}
	\centering
	\caption{Two different point\addra{s} of views for the (same) $3$-dimensional $95\%$-confidence ellipsoid based on the CRLB, with $100$ SBL estimates superimposed (legend as in Fig.~\ref{fig:CRLB_2D_ellipse}, right). The CRLB accurately quantifies how the variance of the proposed solution is spread in the $3$-dimensional space.}
	\label{fig:CRLB_3D_ellipse}\vspace{-0.2cm}
\end{figure}

Figure~\ref{fig:CRLB_2D_ellipse} presents the $2$-dimensional setting under consideration, and the $95\%$ confidence ellipse computed using the CRLB \eqref{CRLBonMSEinequality}, with superimposed estimates $\hup_{\SBL}$ obtained for $100$ different noise recordings. Despite the model mismatch with respect to the noise distribution, a good fit is seen between the empirical results and the predicted theoretical accuracy due to the CRLB. Figure~\ref{fig:CRLB_3D_ellipse} reflects the same fit in the $3$-dimensional space. This not only agrees with our analytical derivation of the bound, but also provides an empirical justification for our stochastic noise model. In this regard, we note further that the hybrid signals we use allow us to essentially isolate the (potential) noise-related model mismatch effects, and test our proposed solution with respect to deviations of this sort only. 

\vspace{-0.6cm}
\subsection{Comparison with GCC-PHAT and MFP3}\label{subsec:simulation_comparison}
\begin{figure*}[t]
	\centering
	\begin{subfigure}[b]{0.245\textwidth}
		\centering
		\includegraphics[width=\textwidth]{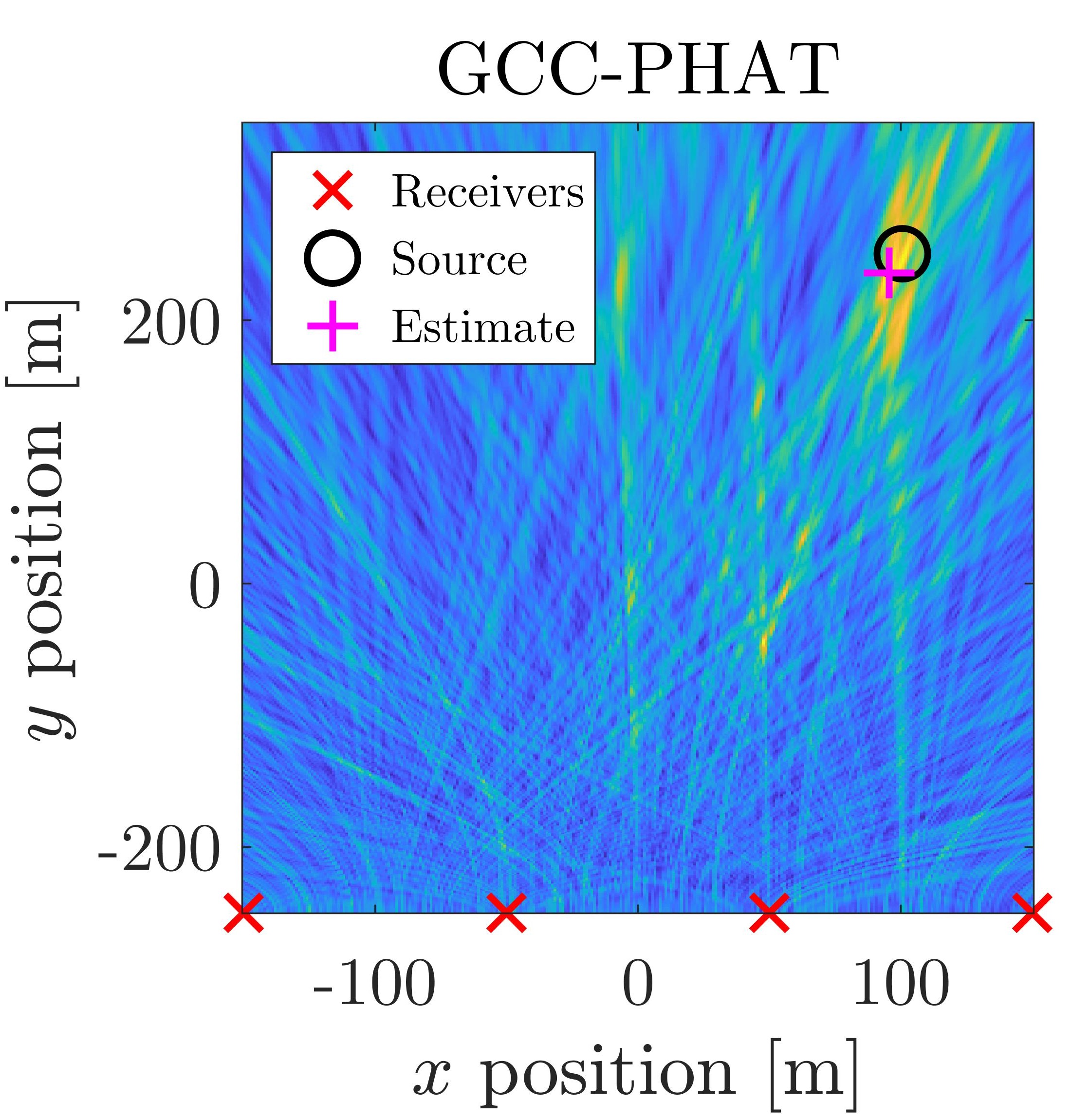}
		\caption{}
		\label{fig:typical_realization_GCCPHAT}
	\end{subfigure}%
	~\hspace{-0.2cm}
	\begin{subfigure}[b]{0.245\textwidth}
		\centering
		\includegraphics[width=\textwidth]{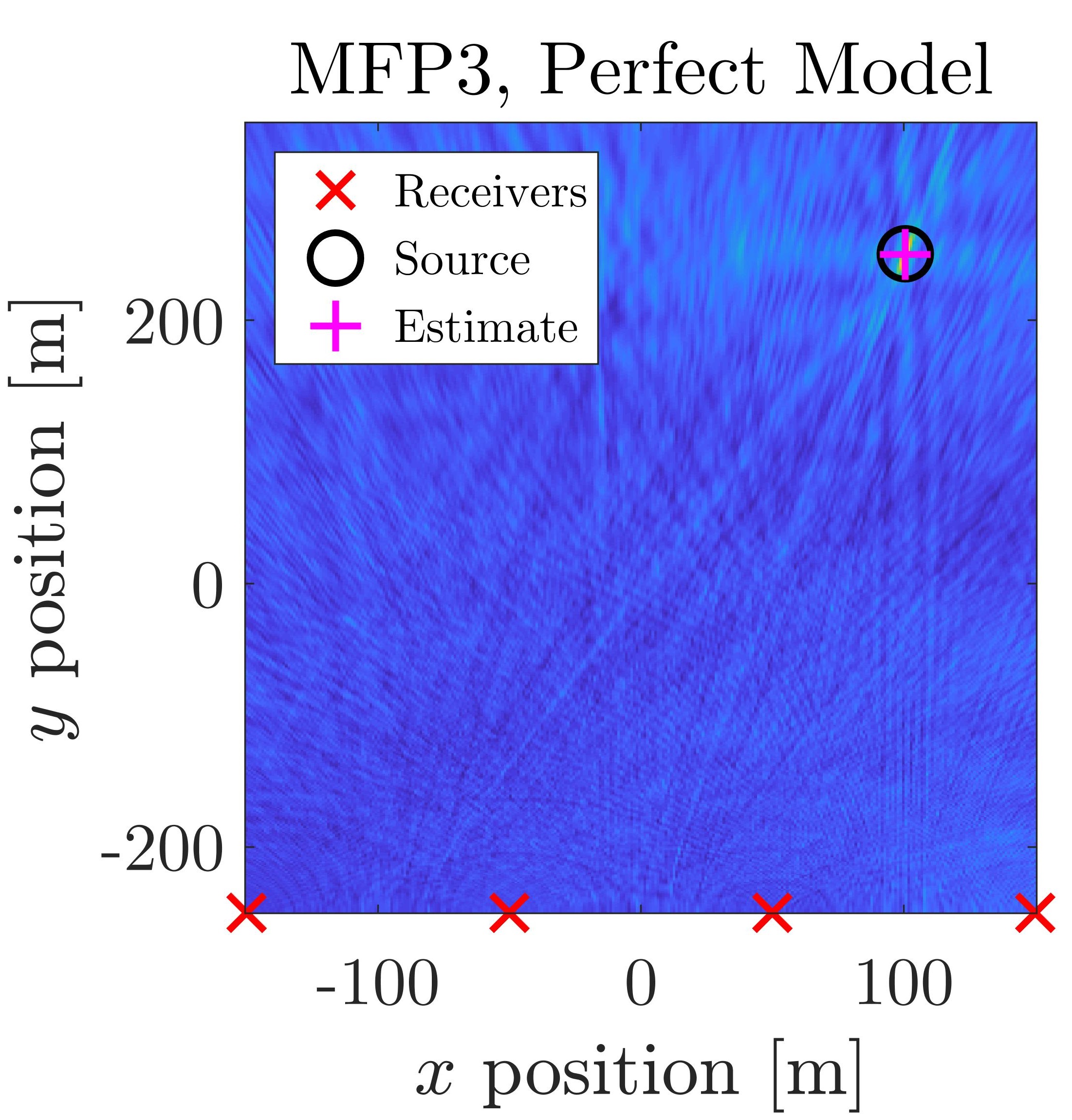}
		\caption{}
		\label{fig:typical_realizations_MFP_perfect}
	\end{subfigure}
	~\hspace{-0.2cm}
	\begin{subfigure}[b]{0.245\textwidth}
		\centering
		\includegraphics[width=\textwidth]{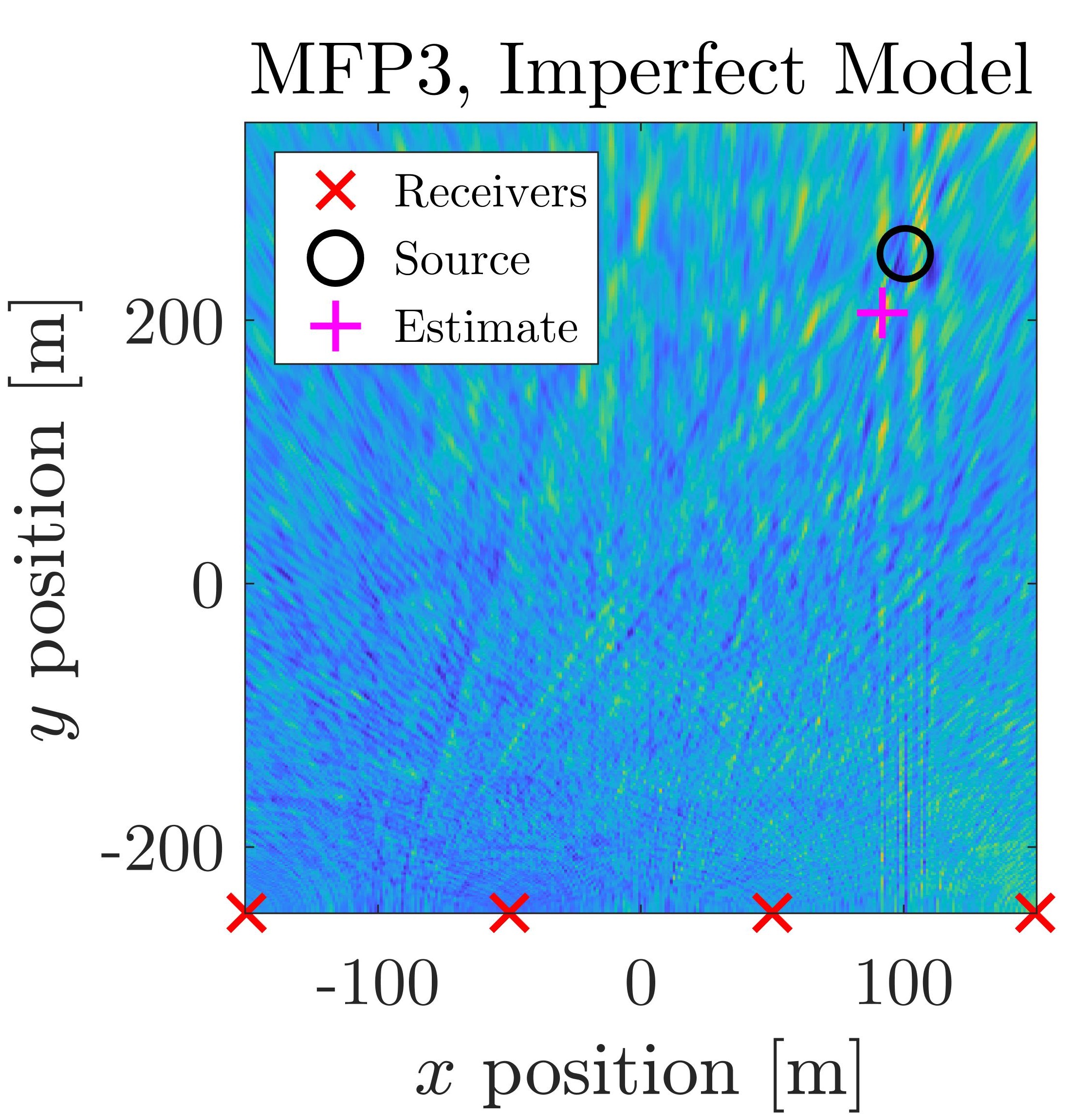}
		\caption{}
		\label{fig:typical_realization_MFP}
	\end{subfigure}
	~\hspace{-0.2cm}
	\begin{subfigure}[b]{0.245\textwidth}
		\centering
		\includegraphics[width=\textwidth]{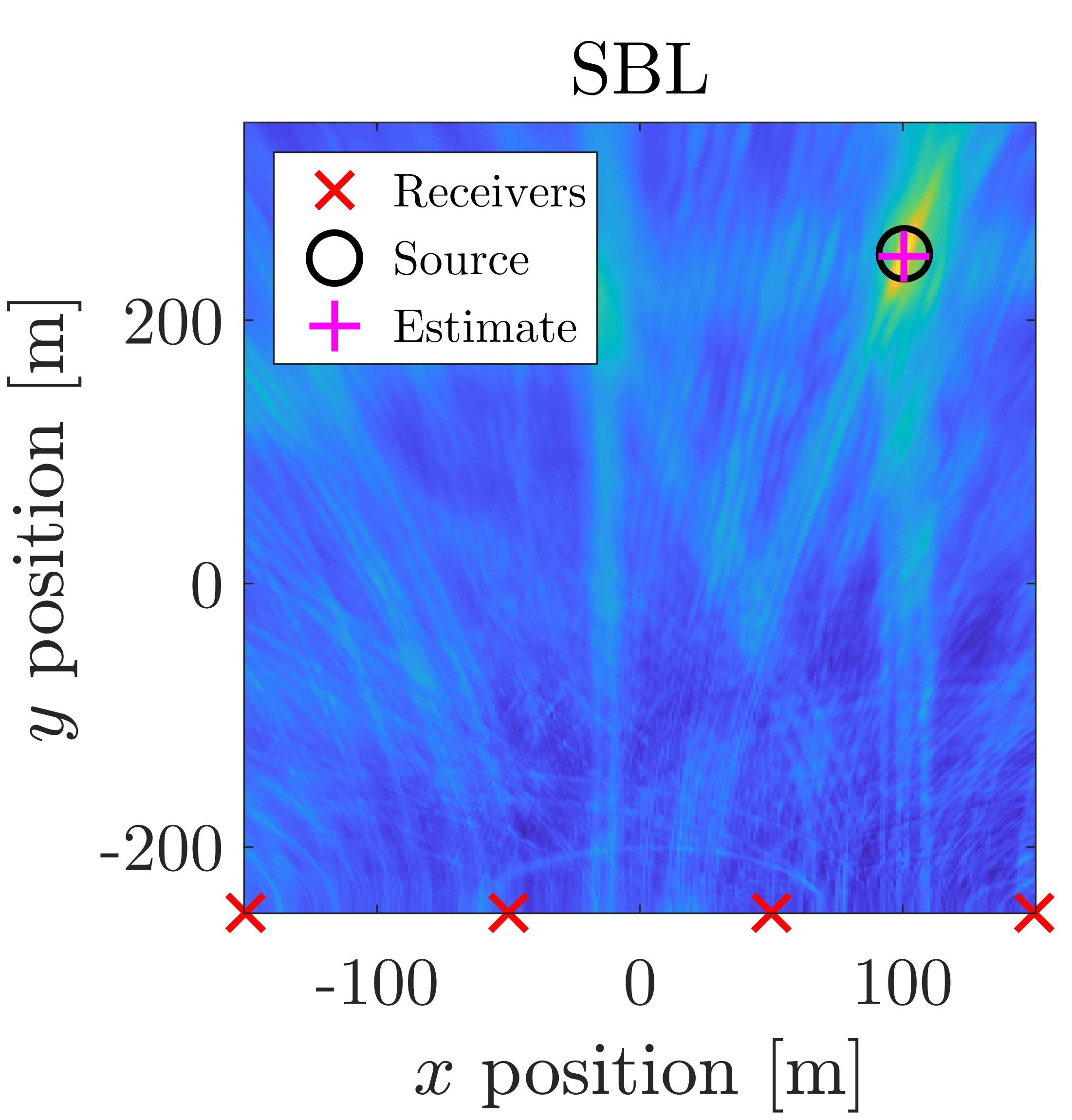}
		\caption{}
		\label{fig:typical_realization_SBL}
	\end{subfigure}
	\caption{$2$-dimensional slices at the source depth ($z=z_p$) of the objective function for typical realizations at $5$ dB SNR. (a) GCC-PHAT (b) MFP3 with perfect model (c) MFP3 with unknown phases of $\ub_{1}\,\ldots,\ub_{L}$ (d) SBL. Evidently, GCC-PHAT and MFP3 without perfect knowledge are considerably more fragile than the proposed method, which is similar to MFP3 with perfect knowledge with respect to stability, at the cost of a higher variance.}
	\label{fig:typical_realizations}
\end{figure*}
\begin{figure*}
\begin{minipage}[t]{0.32\textwidth}
  \includegraphics[width=\linewidth]{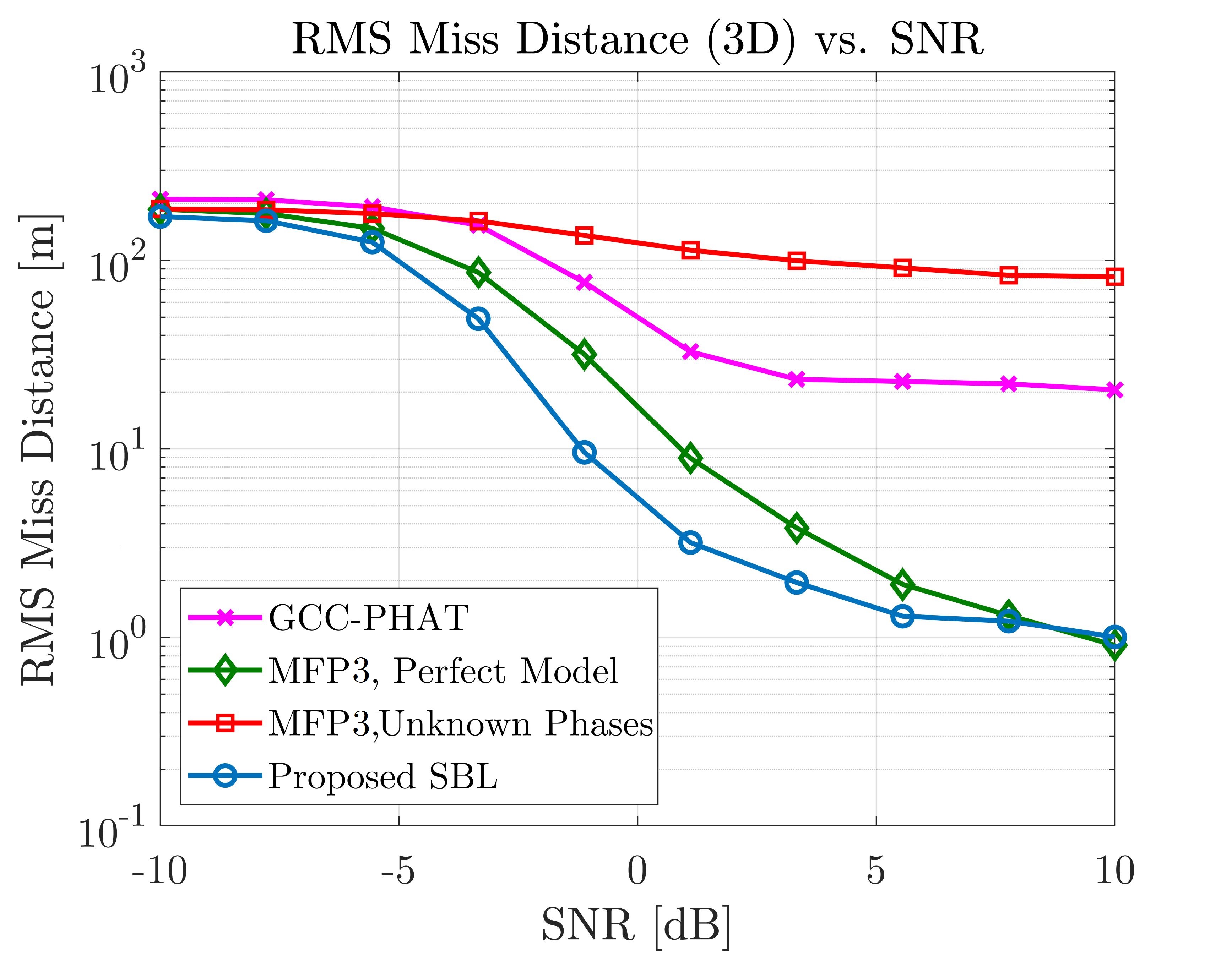}
  \caption{RMS miss distance vs.\ SNR, for $T=100$. At very low SNR, all methods perform poorly. As the SNR increases, GCC-PHAT improves only moderately, while SBL improves significantly.}
  \label{fig:RMSE_vs_SNRexp2}
\end{minipage}%
\hfill 
\begin{minipage}[t]{0.32\textwidth}
  \includegraphics[width=\linewidth]{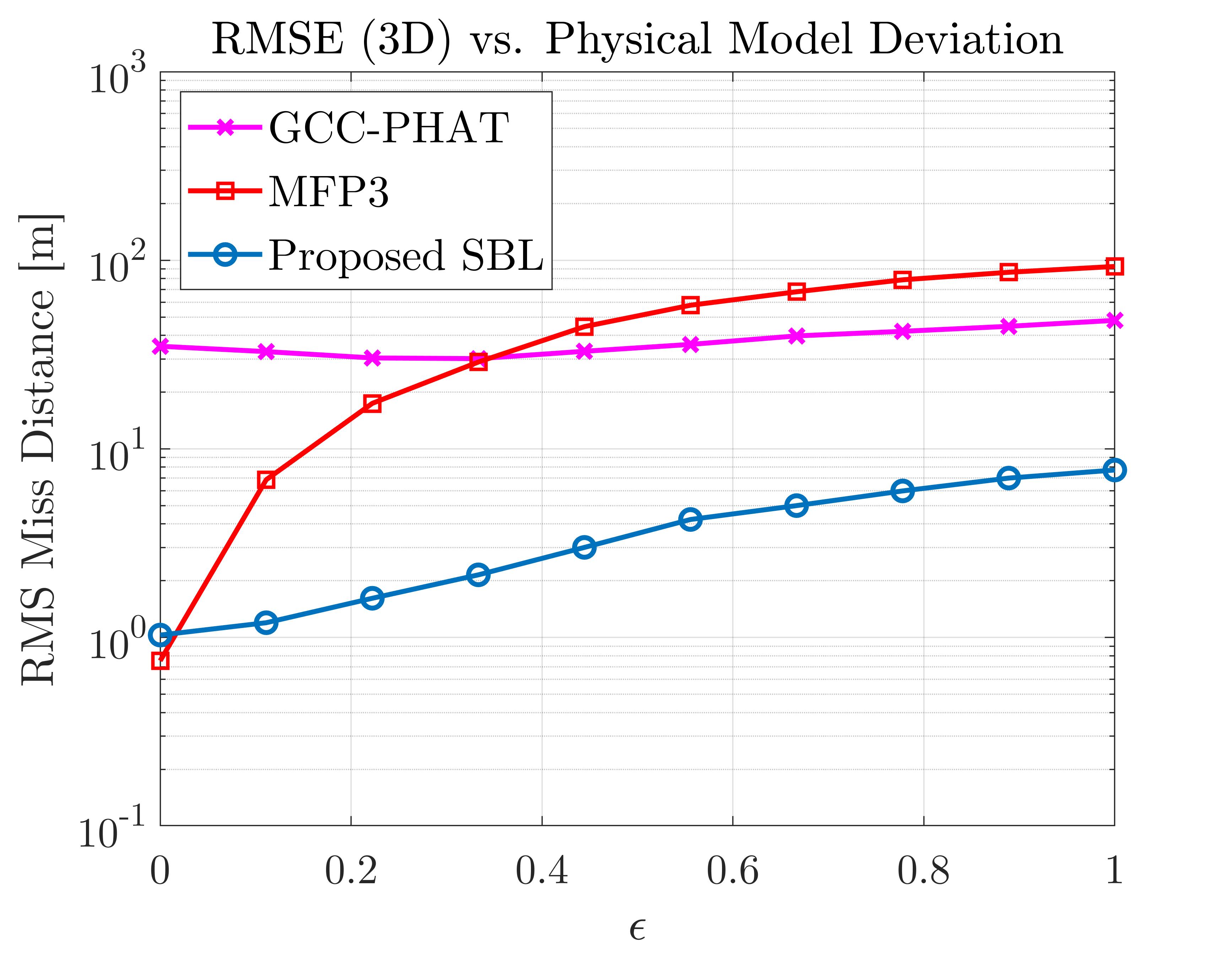}
	\caption{RMS miss distance vs.\ $\epsilon$, quantifying a deviation from the physical model, for $T=100$ and $\sigma_{v_{\ell}}^2=0.1$. MFP3 is sensitive to such model deviations, while GCC-PHAT and SBL are robust.}
	\label{fig:RMSE_vs_epsilon_exp2}
\end{minipage}%
\hfill
\begin{minipage}[t]{0.32\textwidth}
  \includegraphics[width=\linewidth]{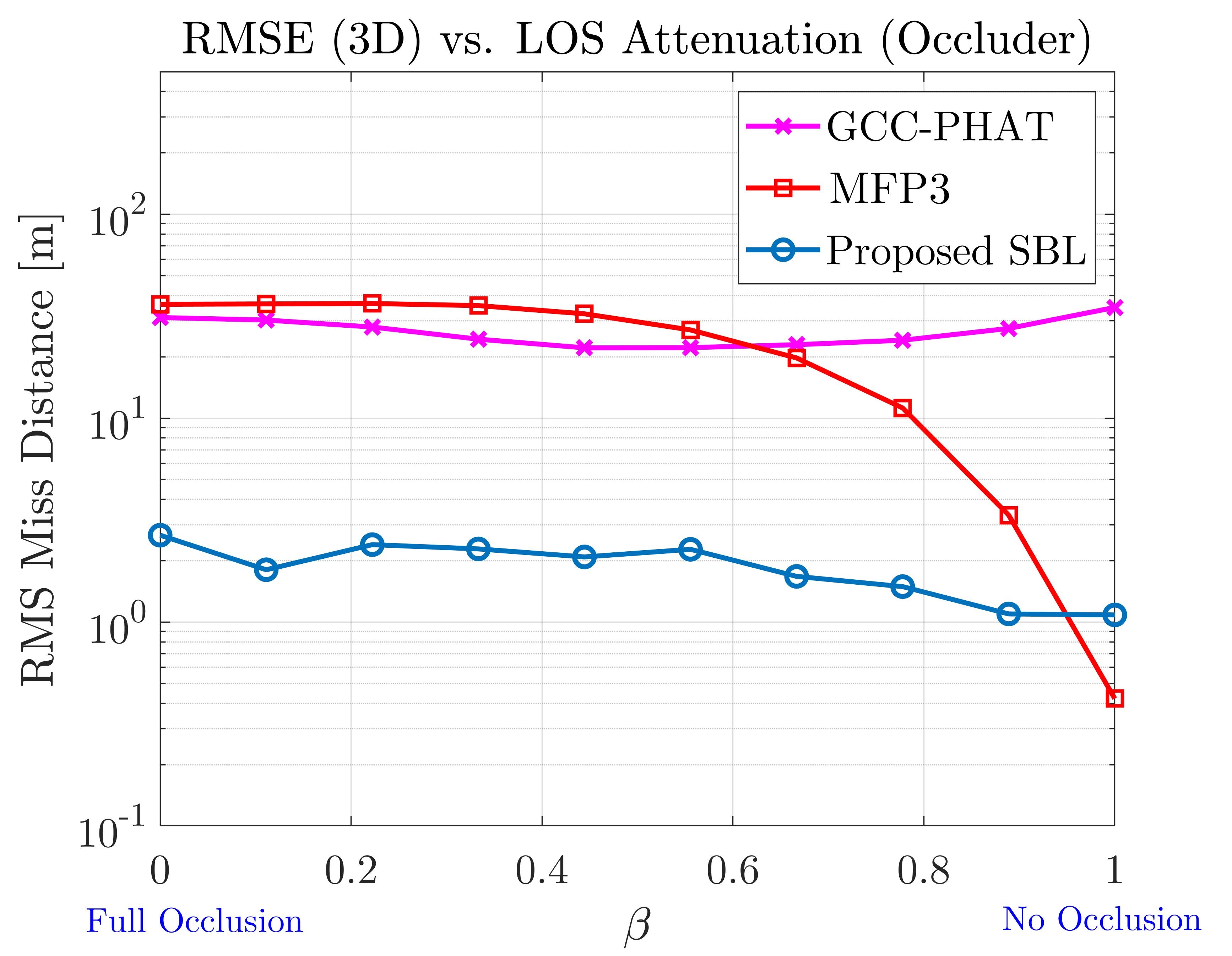}
	\caption{RMS miss distance vs.\ $\beta$, the LOS attenuation coefficient of receivers $\ell=2,3$, for $T=100$ and $\sigma_{v_{\ell}}^2=0.1$. Our proposed SBL offer the best accuracy-stability balance out of the three methods.}
	\label{fig:RMSE_vs_LOS_coeff_exp2}
\end{minipage}%
\vspace{-0.4cm}
\end{figure*}
We now compare the proposed SBL method with MFP3 and the GCC-PHAT localization methods. In this simulated experiment, we consider the setup depicted in Fig.~\ref{fig:typical_realizations}, namely a linear deployment of the receivers. Such a deployment is conceivable for naval defense purposes near the shoreline, or harbor monitoring \cite{van2014underwater}. The positions of the source and the receivers, and all relevant system and environmental parameters are given in Table \ref{table:positionsforExp2}. The source's DFT coefficients $\sbar[k]$, as well as the noise realizations $\{\vbar_{\ell}[k]\}$, were drawn independently from the standard CN distribution in each trial. The noise variance of the $\ell$-th receiver is set as $\sigma_{v_{\ell}}^2=\sigma_v^2\cdot\|\ub_{\ell}\|^2$, and the SNR is defined here as $\Eset\left[|\sbar[k]|^2\right]/\sigma_v^2=1/\sigma_v^2$. All empirical results presented in this subsection are based on averaging $10^4$ independent trials.

\begin{table}
    \centering
	\noindent\makebox[0.515\textwidth]{
	\begingroup
	\setlength{\tabcolsep}{10pt} 
	\renewcommand{\arraystretch}{1.3} 
	\begin{tabular}{|c|c|c|c|l}
		\cline{1-4}
		& $x$ {[}m{]} & $y$ {[}m{]} & $z$ {[}m{]} &  \\ \cline{1-4}
		Source Position, $\up$ &   $100.5976$      &   $250.5837$         &    $30.1131$        &  \\ \cline{1-4}
		Receiver 1, $\up_1$    &     $150$       &    $-250$        &       $10$     &  \\ \cline{1-4}
		Receiver 2, $\up_2$     &     $50$       &    $-250$        &   $15$         &  \\ \cline{1-4}
		Receiver 3, $\up_3$    &    $-50$        &    $-250$        &       $20$     &  \\ \cline{1-4}
		Receiver 4, $\up_4$    &    $-150$        &   $-250$         &       $25$     &  \\ \cline{1-4}
		\multicolumn{4}{|l|}{$c=1535$ m/s, $\kappa_b=0.85$, $h=100$ m, $N=100$, $T_s=0.001$ s}                                                        &  \\ \cline{1-4}
	\end{tabular}
	\endgroup}
	\caption{The setting considered in Subsection \ref{subsec:simulation_comparison}.}
	\label{table:positionsforExp2}
\end{table}
\setlength{\textfloatsep}{2.5pt}

We first compare the localization accuracies of the methods for different SNRs. Figure~\ref{fig:RMSE_vs_SNRexp2} presents the root mean squared (RMS) miss distance, i.e., the square root of \eqref{MSEdefinition}, vs.\ the SNR for each method. A 2D slice at the receiver's depth of the objective functions of each of the algorithms for a typical realization at $5$ dB SNR is given in Fig.~\ref{fig:typical_realizations}. For MFP3, we show the performance obtained with perfect knowledge of the channel (``Perfect Model"), i.e., when $\ub_{1},\ldots,\ub_{L}$ and $\kappa_b$ are known exactly; and when this perfect knowledge is accurate except for the phases of $\ub_{1},\ldots,\ub_{L}$ (``Imperfect Model"), which are drawn independently from $U(0,2\pi)$. As observed, although GCC-PHAT improves when the SNR increases, it essentially cannot cope well---in a $3$-dimensional space optimization---with the addition of the surface and bottom reflections. It is also seen the MFP3 is highly sensitive to deviations from the assumed channel response. In contrast, such deviations are completely transparent to SBL, as it considers these parameters as unknown, and implicitly optimizes over them jointly with all the other unknowns (see \eqref{bsolutionLS}, Appendix \ref{proofofprop1}). The robustness at the moderate cost in performance relative to MFP3 is evident. Note that MFP3's superior performance is guaranteed only asymptotically, in agreement with the results in Fig.~\ref{fig:RMSE_vs_SNRexp2}.

Next, we compare the performances of the three different methods with respect to perturbations in the expected channel attenuations, as prescribed by the physical model \eqref{LOSatten}--\eqref{bottomamplitude}. This form of model mismatch is likely to occur in practice due to non\delra{-}idealities\footnote{Within the three-ray model. Of course, in practice there are more modeling mismatch factors due to the simplified three-ray model. The effects of some of these non\delra{-}idealities will be evaluated in the next experiment, where we apply our method to real data, and demonstrate successful localization.} (e.g., inaccurate prior knowledge of $\kappa_b$). Formally, we model these deviations by generating the channel attenuation coefficients as
\begin{equation}\label{modelmissmatchbz}
\begin{gathered}
b_{r\ell}(\epsilon)=(1-\epsilon\cdot \gamma_{r\ell})b_{r\ell}\cdot e^{\jmath2\pi \epsilon\cdot\varphi_{r\ell}},\\
r\in\{1,2,3\},\; \ell\in\{1,2,3,4\},
\end{gathered}
\end{equation}
where $\{\gamma_{r\ell}\sim U(0,0.5)\}$ and $\{\varphi_{r\ell}\sim U(0,1)\}$ are independent. In \eqref{modelmissmatchbz}, $\epsilon\in[0,1]$ is a parameter controlling the deviation from the physical model, where $\epsilon=0$ corresponds to no deviation from \eqref{LOSatten}--\eqref{bottomamplitude}.

Figure~\ref{fig:RMSE_vs_epsilon_exp2} presents the RMS miss distance vs.\ $\epsilon$. As expected, we observe an overall accuracy-robustness superiority of SBL relative to the competing algorithms. While MFP3 is superior when perfect knowledge of the channel parameters $\kappa_b,\ub_{1},\ldots,\ub_{L}$ is available, SBL is inherently indifferent to deviations from their ideal physical values. GCC-PHAT is also robust to such deviations, but completely ignores (by design) the multipath channel, and therefore cannot exploit additional signal components, such as surface and bottom reflections.

In the last simulation for this setup, we model the effect of a potential occluder between some of the receivers and the source. Specifically, for the second and third receivers (i.e., at $\up_2$ and $\up_3$), we introduce an attenuation coefficient $\beta\in[0,1]$ to the LOS components, such that
\begin{equation}\label{betadefinition}
b_{1\ell}(\beta)=\beta\cdot b_{1\ell}\cdot e^{\jmath2\pi(1-\beta)\varphi_{1\ell}}, \ell\in\{2,3\}.
\end{equation}
When $\beta=1$, there is no occlusion, and when $\beta=0$, the LOS components of receivers $\ell=2,3$ are completely lost. The phase perturbation models the interaction with the occluder.

Figure~\ref{fig:RMSE_vs_LOS_coeff_exp2} presents the RMS miss distance vs.\ $\beta$. It is observed that the accuracy obtained by GCC-PHAT is on the same order of the distances ($\sim25$m) corresponding to the time delays between the LOS and NLOS components. This level of accuracy is stable, but is not satisfactory for an SNR level of $10$ dB (here, $\sigma_v^2=0.1$). It is also seen that in the absence of a modeling error (i.e., $\beta=1$), MFP3 attains the highest accuracy. However, deviations from the ideal signal model, in the form of occluded LOS components of two receivers, inflict a severe performance deterioration. 

As we have demonstrated in Fig.~\ref{fig:typical_realization_MFP} and Fig.~\ref{fig:RMSE_vs_SNRexp2}, when MFP3 is actually mis-matched, it could perform even worse than what is presented in Fig.~\ref{fig:RMSE_vs_LOS_coeff_exp2}. Still, even in this setting, our method exhibits the best accuracy-stability trade-off.

\vspace{-0.3cm}
\subsection{Experimental Results}\label{subsec:HiFAATresults}
\begin{figure}[t]
	\includegraphics[width=0.485\textwidth]{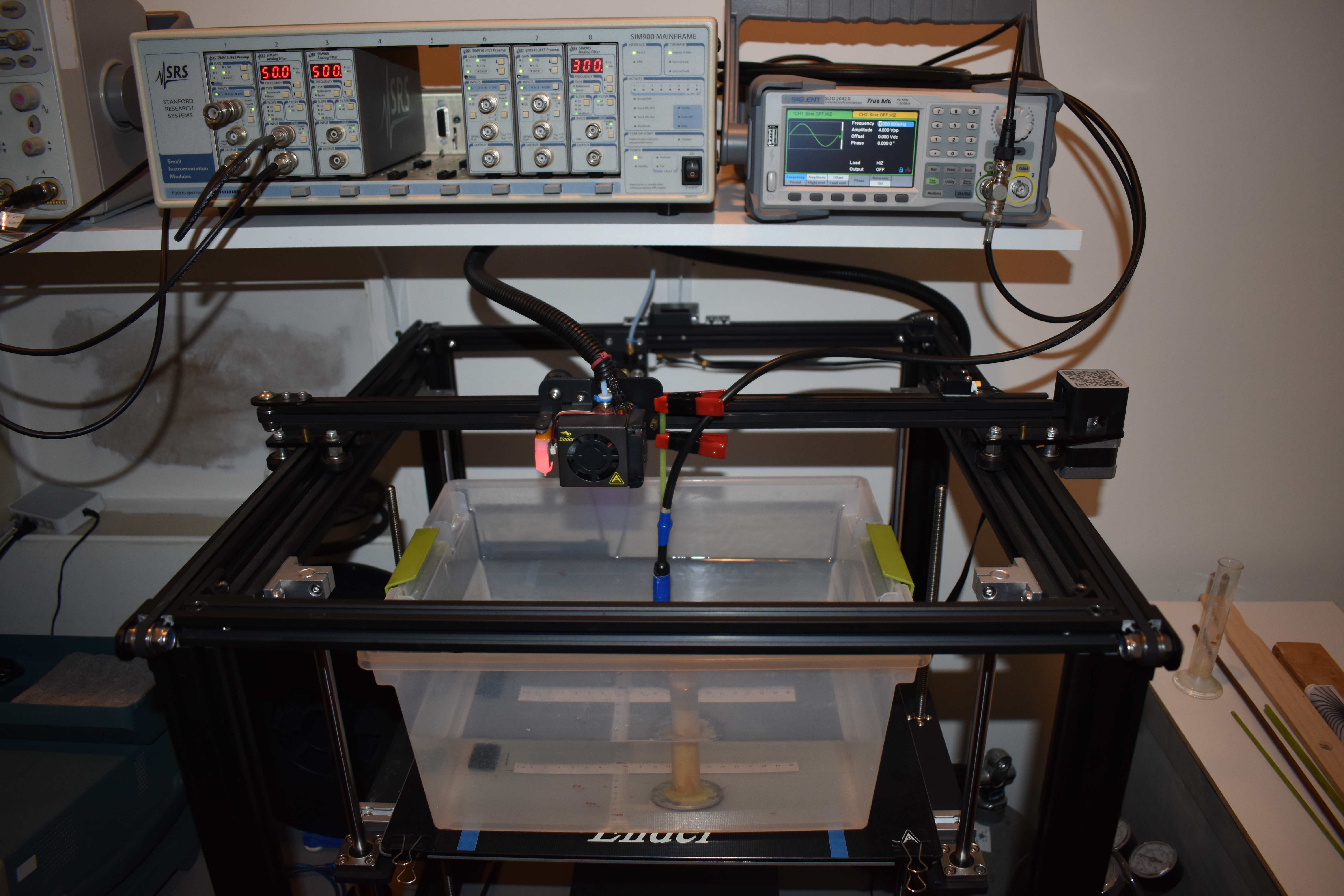}
	\centering
	\caption{A picture of our water tank. Our testbed provides high frequency ($200-400$ kHz) noisy acoustic data from a complex reverberant environment.}
	\label{fig:hifaat_watertank_image}
\end{figure}
We now demonstrate the performance of our proposed method on acoustic data acquired in our water tank testbed---the high frequency autonomous acoustic tank. This system, presented in Fig.~\ref{fig:hifaat_watertank_image}, is (roughly) of size $25\;\text{cm} \times 32\;\text{cm} \times 15\;\text{cm}$, and enables us to create a controlled and challenging setting for frequency-scaled underwater localization.

Although the water tank environment is only a scale model of a shallow-water environment, it nevertheless poses a challenging scenario. In addition to the modeled bottom and surface reflections, the water tank has four additional sides, that are reflective boundaries. These thin plastic boundaries are highly reflective, so that the test environment is highly reverberant, giving rise to a rich multipath channel. In particular, the magnitudes of the unmodeled reflections are comparable to the modeled ones in our three-ray model.

In this experiment, the source is transmitting a Gaussian pulse at a carrier frequency of $280$ kHz, and the speed of sound in the water tank is $c=1485$ $\text{m}/\text{s}$. To maintain consistency across different trials, the source and receivers were set at the same depth. In this case, spatial diversity in the depth-direction is limited, hence we assume here that the source's depth is known, and approach this $2$-dimensional problem.

The received signal\addra{s} w\delra{as}\addra{ere} sampled at $2$ GHz. Before applying the localization method, the signals were decimated by a factor of $2\times10^3$, to obtain \delra{a }$1$ MHz bandwidth signals. For each setting, in which the source and receivers were static, the observation interval was $0.5$ ms long. More technical details are given in the supplementary materials.

\begin{figure*}[t]
	\includegraphics[width=\textwidth]{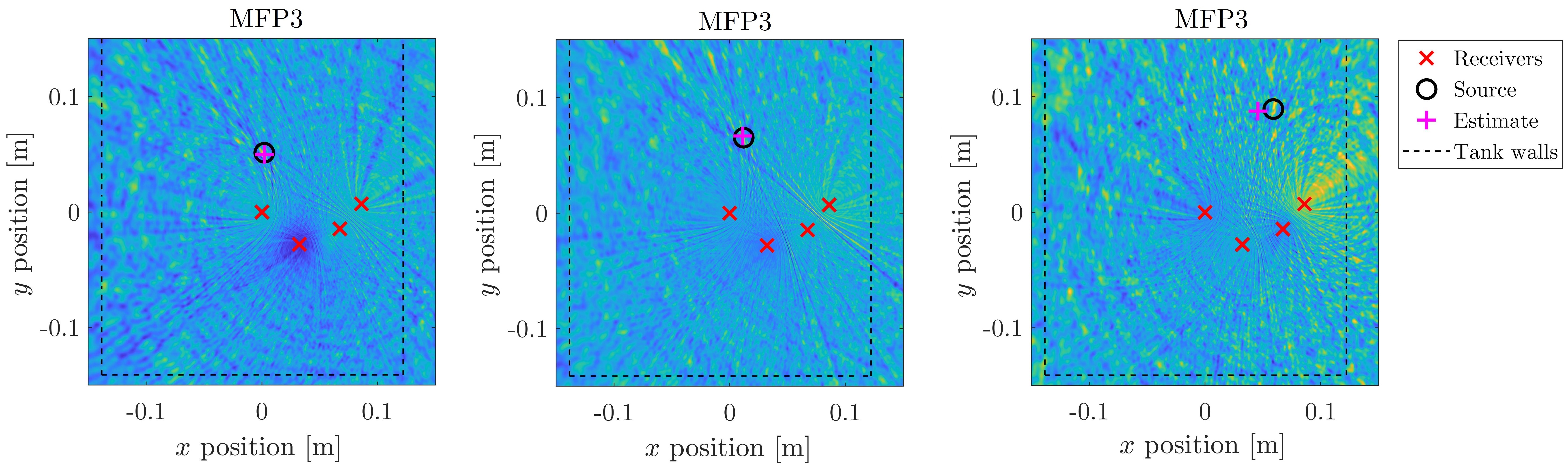}
	\centering
	\caption{MFP3 experimental results based on acoustic measurements acquired in the water tank using the system presented in Fig.~\ref{fig:hifaat_watertank_image}.}
	\label{fig:HiFAAT_MFP_no_cylinder}\vspace{-0.3cm}
\end{figure*}

\begin{figure*}[t]
	\includegraphics[width=\textwidth]{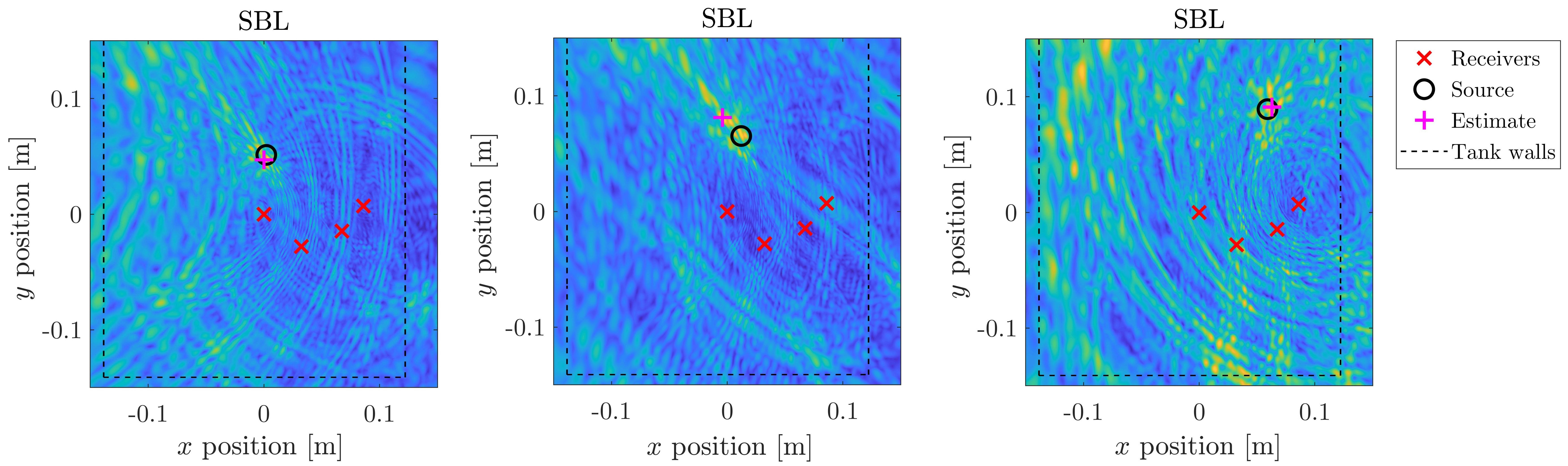}
	\centering
	\caption{SBL experimental results based on acoustic measurements acquired in the water tank using the system presented in Fig.~\ref{fig:hifaat_watertank_image}.}
	\label{fig:HiFAAT_SBL_no_cylinder}
\end{figure*}

Figs.\ \ref{fig:HiFAAT_GCCPHAT_no_cylinder}, \ref{fig:HiFAAT_MFP_no_cylinder} and \ref{fig:HiFAAT_SBL_no_cylinder} present the objective functions of GCC-PHAT, MFP3 and the proposed method, respectively, for three different source locations. Here, the search area is a $30 \times 30\;\text{cm}^2$ square, centered around the first receiver. Note that we intentionally did not align the search area with the one dictated by the boundaries of the tank, since we assume that such prior knowledge is unavailable as in a real problem setting.
\begin{figure*}[t]
	\includegraphics[width=\textwidth]{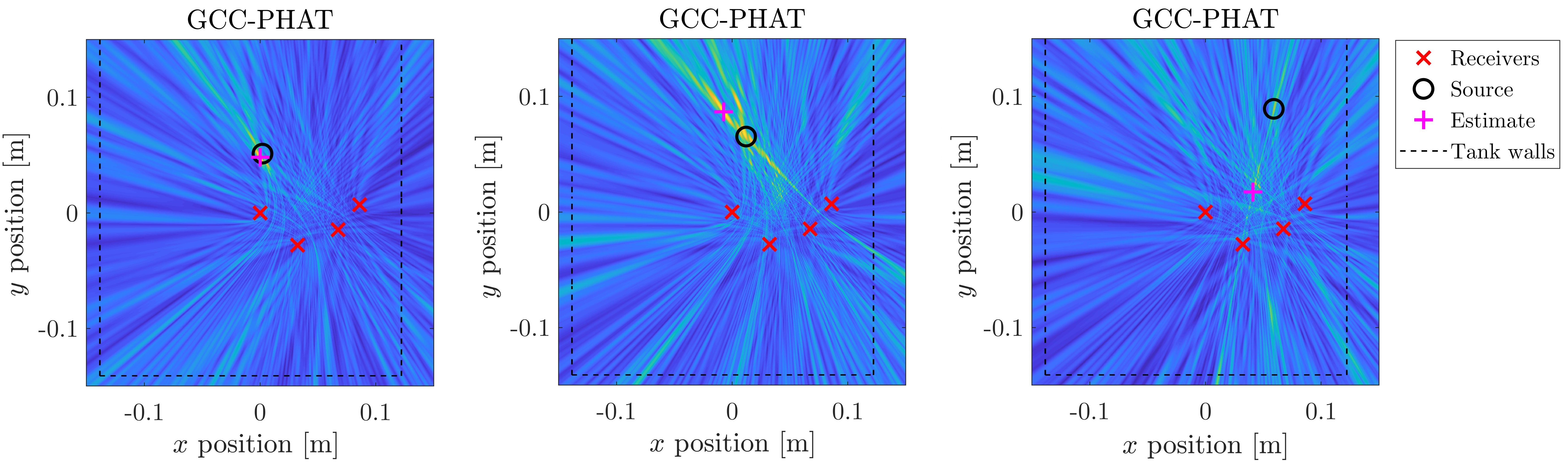}
	\centering
	\caption{GCC-PHAT experimental results based on acoustic measurements acquired in the water tank using the system presented in Fig.~\ref{fig:hifaat_watertank_image}.}
	\label{fig:HiFAAT_GCCPHAT_no_cylinder}\vspace{-0.3cm}
\end{figure*}
As was observed in the simulations, it is seen that in the presence of strong multipath, GCC-PHAT suffers from the worst performance degradation, and MFP3 is the most accurate, best exploiting the environmental prior knowledge. SBL is less accurate than MFP3, but still provides reasonable estimates in the vicinity of the source's true location.

\begin{figure*}[t!]
	\includegraphics[width=\textwidth]{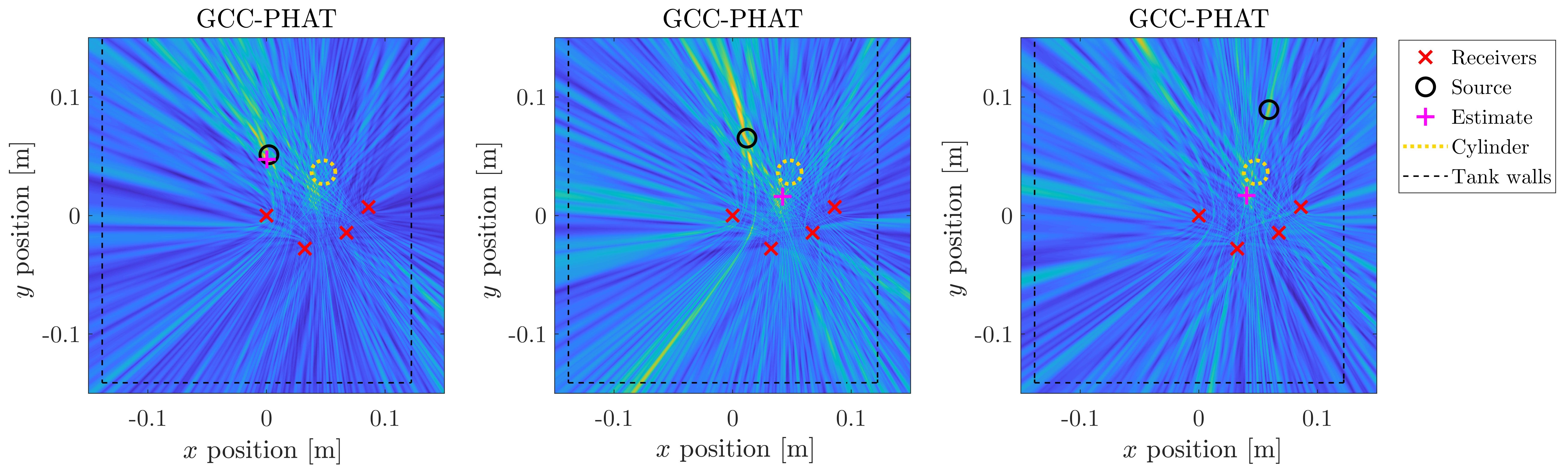}
	\centering
	\caption{GCC-PHAT experimental results based on acoustic measurements acquired in the water tank using the system presented in Fig.~\ref{fig:hifaat_watertank_image}, with the presence of an unknown occluder (cylinder), modeling an effect of inaccurate environmental prior knowledge.}
	\label{fig:HiFAAT_GCCPHAT_with_cylinder}\vspace{-0.3cm}
\end{figure*}

\begin{figure*}[t!]
	\includegraphics[width=\textwidth]{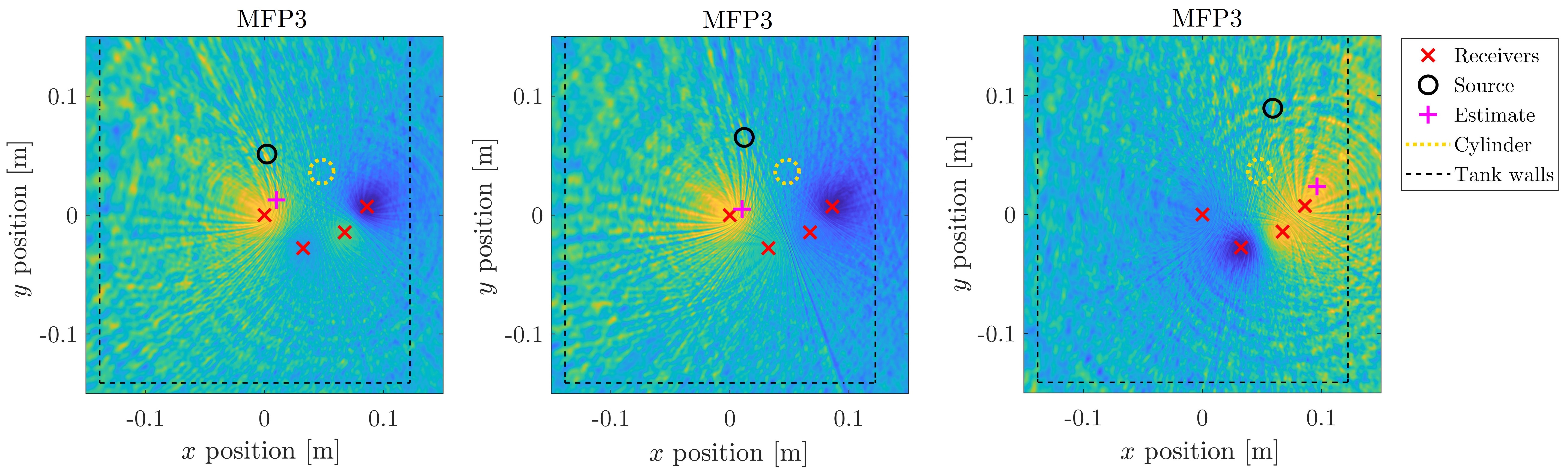}
	\centering
	\caption{MFP3 experimental results based on acoustic measurements acquired in the water tank using the system presented in Fig.~\ref{fig:hifaat_watertank_image}, with the presence of an unknown occluder (cylinder), modeling an effect of inaccurate environmental prior knowledge.}
	\label{fig:HiFAAT_MFP_with_cylinder}\vspace{-0.3cm}
\end{figure*}

\begin{figure*}[t!]
	\includegraphics[width=\textwidth]{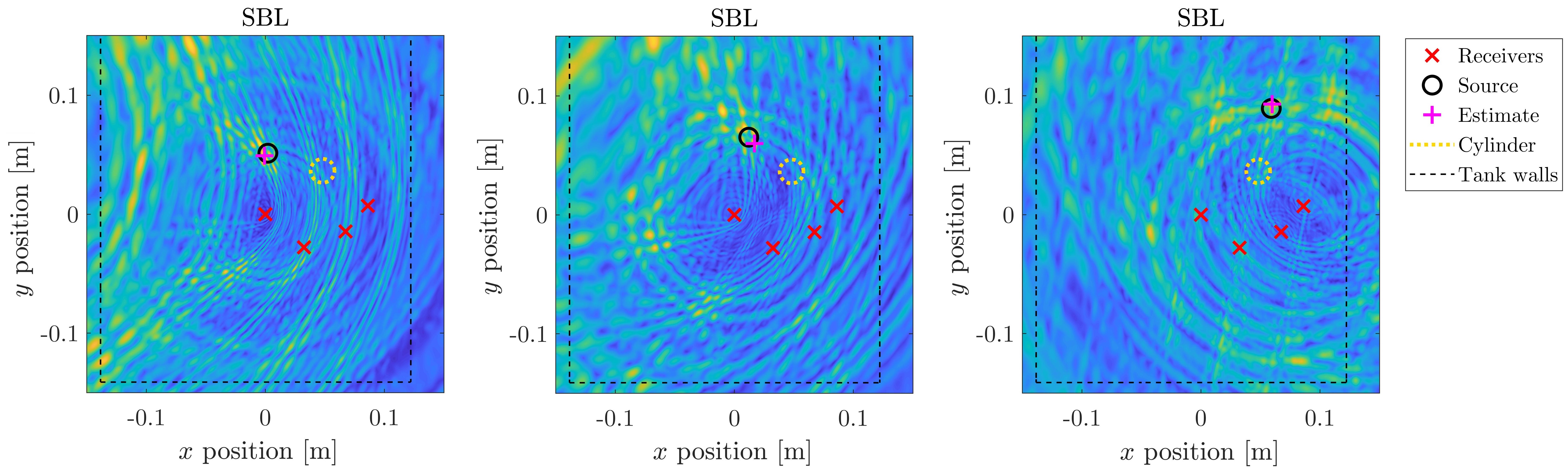}
	\centering
	\caption{SBL experimental results based on acoustic measurements acquired in the water tank using the system presented in Fig.~\ref{fig:hifaat_watertank_image}, with the presence of an unknown occluder (cylinder), modeling an effect of inaccurate environmental prior knowledge.}
	\label{fig:HiFAAT_SBL_with_cylinder}
\end{figure*}

Next, we repeat the experiment but now with the presence of an unknown object---a cylinder, stretching from the bottom to the surface of the tank, placed in the area between the source and the receivers, as depicted in Figs.\ \ref{fig:HiFAAT_GCCPHAT_with_cylinder}--\ref{fig:HiFAAT_SBL_with_cylinder}. This unknown feature causes severe model mismatch; critically, if it blocks the LOS between the source and a receiver, then all the three modeled rays---LOS, and surface and bottom reflections---are essentially blocked. A flexible algorithm can in principle select (possibly implicitly) which receivers to use, and would be able to reject \delra{non-}\addra{un}informative measurements, such as the ones acquired by a receiver ``viewing" the occluded scenery.

Figures~\ref{fig:HiFAAT_GCCPHAT_with_cylinder}, \ref{fig:HiFAAT_MFP_with_cylinder} and \ref{fig:HiFAAT_SBL_with_cylinder}, presenting the objective functions of GCC-PHAT, MFP3 and SBL, respectively, for the same scenarios but with an occluder, corroborate the robustness of the SBL method. It is seen that GCC-PHAT and MFP3 are fragile when such unknown environmental features are present. The SBL method, which exhibits robustness in the presence of the occluder, is still able to localize the source. Since the attenuation coefficients are considered to be unknown, and are implicitly estimated \eqref{bsolutionLS}, a perfectly valid estimated value (for some of them) is a value close or equal to zero. This essentially means that the SBL assigns different weights to measurements from different receivers, thus implicitly choosing to effectively ignore the less informative data acquired by receivers with occluded scenery.


\addra{\section{Discussion and Extensions}\label{sec:multiplesources}
A natural extension of the direct localization problem that we considered, is to the case of multiple sources and/or a more complex channel model, which is beyond the scope of this paper and is left for future work. However, in this section we outline some key challenges of this setting to motivate this non-trivial extended problem. We then discuss the potential use of the SBL estimator in such scenarios, and point out important aspects of identifiability. Before addressing these topics, we first discuss some system design considerations for the particular setting described in Section \ref{sec:problemformulation}.

\subsection{System Design Considerations}
For given, limited resources, one may be interested in enhancing the performance of a system as much as possible with respect to the available degrees of freedom. In the specific case of our localization problem, performance can be understood as accuracy (e.g., in terms of \eqref{MSEdefinition}), and resources, perhaps, as the number of receivers, $L$.\footnote{\addra{which are also the number of sensors in our formulation.}} While we defer the formulation of this notion into a well-defined problem (possibly using some function of the bound \eqref{CRLBonMSEinequality}) to future work, we comment on the important related aspects of the topology and the number of receivers.

Recall that a single receiver, equipped with a single sensor, receives (in general) three signal components, two of which are reflected from the surface and the bottom, which are respectively above and under the receiver. Thus, environmental knowledge regarding the position of the surface and the bottom is equivalent, in some sense, to having additional virtual receivers (e.g., \cite{siderius1997multipath}) above and under the surface and the bottom, respectively. In this respect, a single receiver already provides some vertical spatial diversity. Still, due to the blind nature of the problem, in which the emitted waveform and the channel coefficients are unknown, the information from a single sensor is insufficient for localization. 

However, two sensors already contain six signal components, and can in principle contain sufficient information for localization. Intuitively, breaking the symmetry ``as much as possible" relative to the environment in which the system is deployed would lead to better performance. In the two receivers case, placing the second receiver in a different horizontal location than the first would lead to an increased horizontal spatial diversity, which would in turn lead to enhanced performance. The principle of increasing spatial diversity with a fixed resources allocation can be formulated in some settings (e.g., \cite{shulkind2018sensor}), and provides guidance and intuition for the design of a sensors network spatial distribution. For example, one may consider deploying a linear network of sensors \emph{obliquely} relative to the surface.

\vspace{-0.2cm}
\subsection{Key Challenges in Extended Models}\label{subsec:keychallenges}
Incorporating multiple sources into the 3-ray signal model \eqref{modelequation} changes the interplay between the (consequently increased) number of unknown parameters. As a result, it is no longer clear whether a simplified, efficiently computable expression for the objective function---as \eqref{SBLestimate} in Proposition \ref{maxeigenvalueflatspectrum}---can be obtained. Recall that this has a significant effect on the overall computational complexity of the method.

Moreover, assume that there are $M$ sources to be localized (where $M$ is known), and further assume that we have obtained such a simplified, computationally efficient expression for the objective function, which is a function of the sources' positions \emph{only}, denoted by, say, $\up^{(1)},\ldots,\up^{(M)}\in\Rset^{3\times1}$. At this point, in order to obtain the optimal direct localization solution (in the sense of the extended criterion of \eqref{MLEoptimmization}) for all $M$ sources, one is required to solve a $3M$-dimensional nonlinear optimization problem, which may well be non-convex. Consequently, even for $M=2$ sources, this is already difficult with a naive extension of our current proposed method, as it would require a $6$-dimensional grid search (referring to the first step of the proposed solution), which is infeasible for reasonable resolutions. Hence, a different approach is perhaps required in order to solve the multiple sources direct localization problem.

Focusing again on the single source case, one may consider an extended $K$-ray model (with $K>3$), assuming it would \emph{accurately} describe the signal propagation, such that the $K$ rays include primary \emph{and} second- and higher-order reflections. In that case, the performance (i.e., accuracy) improvement would be due to an increased effective/post-processing SNR. This can be understood from the interpretation given in Subsection \ref{subsec:sblinterpretation}, where the SBL method is seen as an implicit way to coherently add all the $K$ reflections from all $L$ sensors.

However, and since the ray-based propagation model is an approximation, while the deviations from the $3$ primary rays can be small, the aggregated approximation errors in the time-delays of the higher-order reflections are likely to no longer be negligible. In that case, on top of additional computational burden, a naive extension of the current approach might yield a more fragile estimator, which is sensitive to model mismatch. The challenging task of exploiting more complex propagation-related phenomena for enhanced, computationally attractive direct localization remains to be explored in future work.

\subsection{SBL as a Solution for Multiple Sources}\label{subsec:SBLassuboptimalsolution}
Notwithstanding the above, our proposed algorithm can still be used for multiple sources localization as a sub-optimal, yet computationally feasible solution. Indeed, \eqref{SBLestimate} can be viewed as a spatial quasi-likelihood map (as a function of $\up$), whose $M$ highest maxima correspond to the $M$ points in space, where sources are most likely to be present (under the mismatched model \eqref{modelequation}, treating, for each source, all the other $M-1$ sources as additive noise). While providing analytical guarantees for this case is beyond the scope of the current work, using the Matlab package provided in the supplementary material, one could easily verify that the SBL method still serves as a viable localization solution for this extended setting.

Given any set of parameters that describe a particular localization problem (i.e., bottom depth, locations of the receivers, etc.), the model \eqref{modelequation} is guaranteed to be identifiable when the FIM is nonsingular, namely $\det\left(\J(\utheta)\right)\neq0$, and \eqref{CRLBonMSEinequality} is finite. However, when using the SBL for localization of multiple source (i.e., under mismatched model) as described above, this is obviously no longer true. Indeed, as least theoretically, there are certain ``special" (however somewhat extreme) scenarios in which the sources not only could not be localized (reliably, or at all), but may also ``disappear" from the resulting heatmap.

To illustrate this, consider the following case, which is depicted in Fig.~\ref{fig:twosourceandoccluder}. Assume, for example, that $L$ receivers are all deployed at the same depth $h/2$ in a linear structure, namely $z_{\ell}=z_0=h/2, y_{\ell}=y_0$ and $x_{\ell}=x_0+\Delta(\ell-1)$ for all $\ell\in\{1,\ldots,L\}$, where $\Delta$ is the spacing between the receivers. Now, further assume that \emph{two} sources are present, such that the second source is located at the same horizontal location as the first, but is located symmetrically about the half depth $h/2$ relative to the first. That is, if the first source is at $\up=(x_p,y_p,z_p)$, the second is at $\widetilde{\up}=(x_p,y_p,h-z_p)$. Finally, assume that an occluding object is present, such that (only) all LOS components at all $L$ receivers are blocked from both of the sources. Denoting by $\widetilde{R}_{2\ell}, \widetilde{R}_{3\ell}$ the distances traveled by the NLOS surface and bottom associated rays, respectively, from the second source, it readily follows from \eqref{NLOSsurfacedistance}--\eqref{timedelyaformula} that $\widetilde{R}_{2\ell}=R_{3\ell}, \widetilde{R}_{3\ell}=R_{2\ell}$, hence
\begin{equation*}
    \tau_{2\ell}(\up)=\tau_{3\ell}(\widetilde{\up}), \quad \tau_{3\ell}(\up)=\tau_{2\ell}(\widetilde{\up}).
\end{equation*}
In this case, if $\kappa_b=1$ (of \eqref{bottomamplitude}), and if the two sources are collaborating and coordinated, then by transmitting the same waveform, they are essentially ``acoustically invisible" (under the three ray model). Indeed, if we denote the waveform of the second source by $\widetilde{s}(t)=s(t)$, the baseband-converted signal from the $\ell$-th receiver (as in \eqref{modelequation}) would then be
\begin{align*}\label{acousticinvisibility}
    x_{\ell}[n]=&\sum_{r=2}^{3}b_{r\ell}\left.s\left(t-\tau_{r\ell}(\up)\right)\right\vert_{t=nT_s}+\\
    &\underbrace{\sum_{r=2}^{3}\widetilde{b}_{r\ell}\left.\widetilde{s}\left(t-\tau_{r\ell}(\widetilde{\up})\right)\right\vert_{t=nT_s}}_{=\,-\sum_{r=2}^{3}b_{r\ell}\left.s\left(t-\tau_{r\ell}(\text{\boldmath$p$})\right)\right\vert_{t=nT_s}}+v_{\ell}[n]=v_{\ell}[n],
\end{align*}
since $\widetilde{b}_{2\ell}=\frac{-1}{R_{3\ell}}=-b_{3\ell}$ and $\widetilde{b}_{3\ell}=\frac{1}{R_{2\ell}}=-b_{2\ell}$, and we recall that due to the occluder, $b_{1\ell}=\widetilde{b}_{1\ell}=0$ for all $\ell\in\{1,\ldots,L\}$. Thus, only noise is observed, and all the information is lost.

\begin{figure}[t]
	\includegraphics[width=0.48\textwidth]{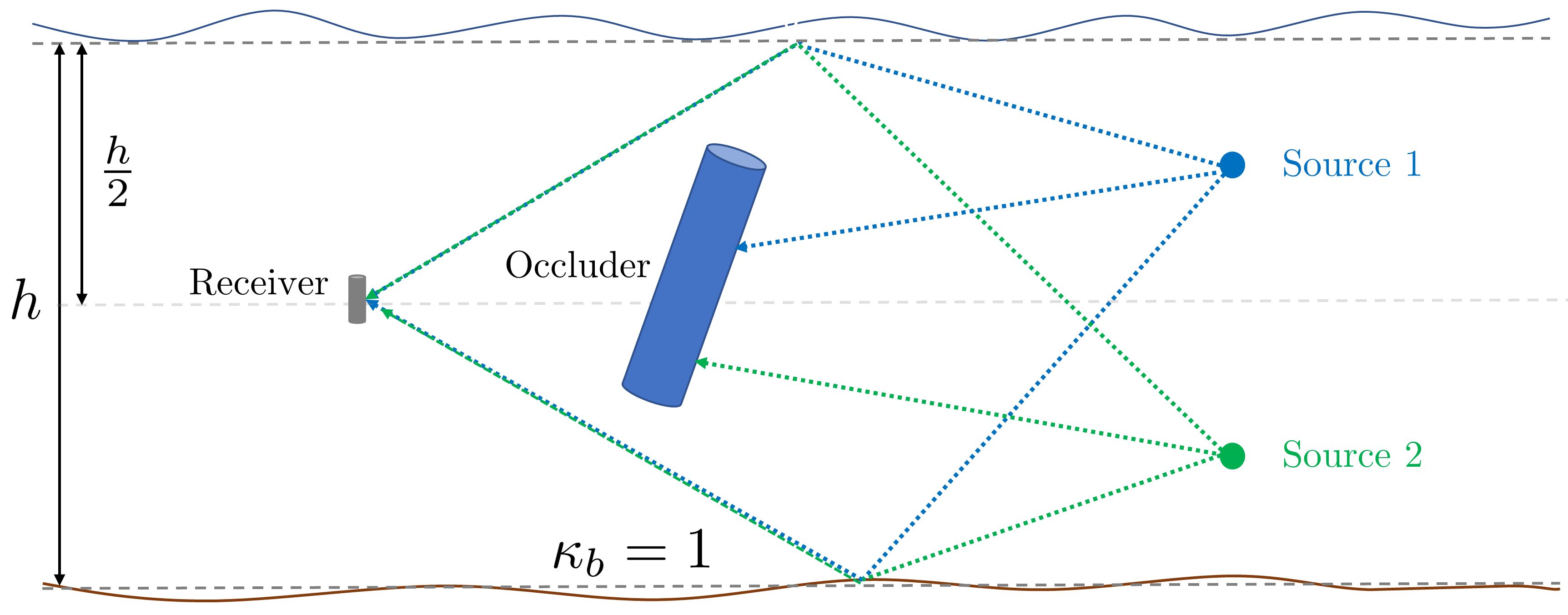}
	\centering
	\caption{\addra{A $2$-dimensional illustration of the special case described in Subsection \ref{subsec:SBLassuboptimalsolution}, wherein the sources are ``acoustically invisible" in terms of the three-ray model, and cannot be localized by the SBL estimator.}}
	\label{fig:twosourceandoccluder}
\end{figure}
While the scenario above can certainly inspire underwater acoustic warfare techniques devised against single-source methods like the SBL, it nonetheless describes an extreme case of a perfectly tailored setting, where several conditions, which are exceptionally difficult to ensure, are fulfilled simultaneously. Therefore, and while this is only one example of a potential failure mechanism of the proposed method when used for localization of multiple sources, one may still gain a general impression of what should happen in order for SBL to completely fail in this setting. Generally, when the sources are not collaborating/coordinated and/or the symmetry is broken (e.g., by a non-regular deployment of the receivers), it is reasonable that the proposed method would still provide reliable localization for multiple sources.
}

\section{\delra{Conclusion}\addra{Concluding remarks}}\label{sec:conclusion}
In the general context of underwater acoustics, based on the three-ray propagation model, we presented a semi-blind localization method, which incorporates environmental knowledge. A closed-form expression for the objective function was derived, along with an equivalent expression, which is more computationally appealing. Thanks to additional degrees of freedom in our model, the proposed method is more robust than its MFP counterpart, and can successfully localize a source in the absence of LOS components. Further, it exhibits stable performance enhancement with respect to methods modeling only LOS components, such as GCC-PHAT.

Since the proposed method is able to coherently ``collect" three signal components from each sensor, the post-processing SNR is consequently higher than any LOS-based method, which collect only one. This way, a given level of localization accuracy can generally be attained with shorter observation intervals. In turn, it is easier to incorporate the SBL method within an appropriate tracking algorithm (such as Kalman filtering), allowing for a more general framework that localizes the source and tracks its movement, assuming the source's velocity is sufficiently \delra{small}\addra{low}. As a topic for future research, in such cases it may be possible to develop a computationally efficient update scheme for the objective function \eqref{maxeigenval}, based on eigenvalue perturbation theory. Another direction for future research, that is of great practical interest, is to apply coarse quantization to the collected data \cite{weiss2021one}, thus reducing the required bandwidth for communication between the different \delra{sensors}\addra{receivers}.\addra{ The above is also true for the potential extensions for multiple sources, and for extended propagation models.}
\addra{
\section{Acknowledgement}
The authors are grateful to James Preisig for helpful discussions, and for providing the KAM11 recordings.
}
\appendices

\section{Proof of Proposition \ref{maxeigenvalueflatspectrum}}\label{proofofprop1}
\begin{proof}
Using the identity $\Diag\left(\D_{\ell}(\up)\ub_{\ell}\right)\bus=\bS\D_{\ell}\ub_{\ell}$, it is easily seen that, for every $\ell\in\{1,\ldots,L\}$, $\widetilde{C}_{\SBL}(\up,\bus,\B)$ of \eqref{NLSobjectivefunction} is minimized with respect to $\ub_{\ell}$ by
\begin{equation}\label{bsolutionLS}
\widehat{\ub}_{\ell}=\left(\left(\bS\D_{\ell}\right)^{\her}\bS\D_{\ell}\right)^{-1}\left(\bS\D_{\ell}\right)^{\her}\bux_{\ell}\delra{.}\addra{,}
\end{equation}
\addra{assuming\footnote{We ignore the extreme, unrealistic cases in which $\D_{\ell}$ are not full rank, which occur only for very specific settings of the receivers' and source's positions. \delra{At any rate}\addra{Nonetheless}, the initial optimization is performed via a grid search, hence we can discard points giving rise to these rare, singular settings.} $\text{rank}\left(\D_{\ell}\right)=3$ for all $\ell\in\{1,\ldots,L\}$ hereafter. }Substituting  $\hB\triangleq[\widehat{\ub}_{1} \cdots \widehat{\ub}_{L}]$ into $\widetilde{C}(\up,\bus,\B)$ yields

\begin{align}
\breve{C}(\up,\bus)_{\SBL}
&\triangleq\widetilde{C}_{\SBL}(\up,\bus,\hB)=\sum_{\ell=1}^{L}\left\|\bux_{\ell}-\bS\D_{\ell}\widehat{\ub}_{\ell}\right\|_2^2\notag\\
&=\sum_{\ell=1}^{L}\big[\bux_{\ell}^{\her}\bux_{\ell}-\bux_{\ell}^{\her}\bS\D_{\ell}\widehat{\ub}_{\ell}-\widehat{\ub}_{\ell}^{\her}\left(\bS\D_{\ell}\right)^{\her}\bux_{\ell} \notag\\
&\qquad\qquad\qquad\qquad{}+\widehat{\ub}_{\ell}^{\her}\left(\bS\D_{\ell}\right)^{\her}\bS\D_{\ell}\widehat{\ub}_{\ell}\big].
\label{Costofsandp}
\end{align}
From \eqref{bsolutionLS}, we observe that
\begin{equation*}
\widehat{\ub}_{\ell}^{\her}\left(\bS\D_{\ell}\right)^{\her}\bS\D_{\ell}\widehat{\ub}_{\ell}=\widehat{\ub}_{\ell}^{\her}\left(\bS\D_{\ell}\right)^{\her}\bux_{\ell},
\end{equation*}
with which \eqref{Costofsandp} simplifies to
\begin{equation}\label{Costofsandpsimple}
\breve{C}_{\SBL}(\up,\bus)=\underbrace{\sum_{\ell=1}^{L}\bux_{\ell}^{\her}\bux_{\ell}}_{\substack{\text{constant with respect to } \\ {\text{\boldmath$p$ and \boldmath$\bar{s}$}}}}-\sum_{\ell=1}^{L}\bux_{\ell}^{\her}\bS\D_{\ell}\widehat{\ub}_{\ell}.
\end{equation}
Therefore, using \eqref{Costofsandpsimple}, \eqref{MLEoptimmization} can now be written as
\begin{equation}\label{simeplmaxobjective}
\min_{\substack{\text{\boldmath$\bar{s}$}\in\mathcal{S}_N \\ {\text{\boldmath$B$}}\in\Cset^{3\times L}}} \widetilde{C}_{\SBL}(\up,\bus,\B) = \max_{\text{\boldmath$\bar{s}$}\in\mathcal{S}_N} \;\sum_{\ell=1}^{L}\bux_{\ell}^{\her}\bS\D_{\ell}\widehat{\ub}_{\ell}.
\end{equation}
At this point, notice that using 
\begin{equation*}
\bux_{\ell}^{\her}\bS=\bus^{\tps}\bX_{\ell}^{*} \Longrightarrow \left(\bS\D_{\ell}\right)^{\her}\bux_{\ell}=\left(\bX_{\ell}^{\her}\D_{\ell}\right)^{\her}\bus^*,
\end{equation*}
we may write
\begin{equation}\label{differentformofbest}
\widehat{\ub}_{\ell}=\left(\D_{\ell}^{\her}\bS^{\her}\bS\D_{\ell}\right)^{-1}\D_{\ell}^{\her}\bX_{\ell}\bus^*.
\end{equation}
Substituting $\widehat{\ub}_{\ell}$ from \eqref{differentformofbest} into \eqref{simeplmaxobjective}, using \eqref{differentformofbest}, $\bS^{\her}\bS=\Diag(|\bus|^2)\triangleq\P_{\sbar}$ and simplifying further yields

\begin{align}
&\max_{\text{\boldmath$\bar{s}$}\in\mathcal{S}_N} \sum_{\ell=1}^{L}\bux_{\ell}^{\her}\bS\D_{\ell}\widehat{\ub}_{\ell}=\nonumber\\
&\max_{\text{\boldmath$\bar{s}$}\in\mathcal{S}_N} \sum_{\ell=1}^{L}\bus^{\tps}\bX_{\ell}^{*}\D_{\ell}\left({\D_{\ell}}^{\her}\P_{\sbar}\D_{\ell}\right)^{-1}\D_{\ell}^{\her}\bX_{\ell}\bus^*=\label{transition1cost}\\
&\max_{\text{\boldmath$\bar{s}$}\in\mathcal{S}_N} \bus^{\her}\left(\sum_{\ell=1}^{L}\bX_{\ell}\D^*_{\ell}\left({\D_{\ell}}^{\tps}\P_{\sbar}\D^*_{\ell}\right)^{-1}\left(\bX_{\ell}\D^*_{\ell}\right)^{\her}\right)\bus\label{Postivecost},
\end{align}
where from \eqref{transition1cost} to \eqref{Postivecost} we have used that $\P_{\sbar}\in\Rset_+^{N\times N}$, and that \eqref{transition1cost} is real-valued (and non\delra{-}negative).

By assumption, $\meps=\mathbf{O}$, hence $\P_{\sbar}=P_s\cdot\I_N$ from \eqref{approxflatsignal}. Thus, in this case \eqref{Postivecost} simplifies further to
\begin{equation*}
\begin{aligned}
&\max_{\text{\boldmath$\bar{s}$}\in\mathcal{S}_N} \bus^{\her}\left(\sum_{\ell=1}^{L}\bX_{\ell}\D^*_{\ell}\left(\D_{\ell}^{\tps}\D^*_{\ell}\right)^{-1}\left(\bX_{\ell}\D^*_{\ell}\right)^{\her}\right)\bus=\\
&\max_{\text{\boldmath$\bar{s}$}\in\mathcal{S}_N} \bus^{\her}\Q(\up)\bus=\lambda_{\max}\left(\Q(\up)\right),
\end{aligned}
\end{equation*}
where $\Q(\up)$ is defined in \eqref{targetmatrixdef}. Therefore, we conclude that when $\meps=\mathbf{O}$, the SBL position estimate is given by
\begin{equation*}
\hup_{\SBL}= \underset{\text{\boldmath$p$}\in\Rset^{3\times1}}{\argmax} \; \lambda_{\max}\left(\Q(\up)\right).
\end{equation*}
\end{proof}

\section{Proof of Proposition \ref{maxeigenvaluenonflatspectrum}}\label{proofofprop2}
\begin{proof}
Observe that in the proof of Proposition \ref{proofofprop1}, \eqref{Postivecost} holds for the general case, where $\meps$ is not necessarily equal to $\mathbf{O}$. Therefore, starting from \eqref{Postivecost}, and focusing on the inverse matrix of a single matrix element in the sum, we now have
\begin{equation*}
\begin{aligned}
&\left(\D_{\ell}^{\tps}\P_{\sbar}\D^*_{\ell}\right)^{-1}=\frac{1}{P_s}\left(\D_{\ell}^{\tps}\left(\I_N + \meps\right)\D^*_{\ell}\right)^{-1}=\\
&\frac{1}{P_s}\left(\D_{\ell}^{\tps}\D^*_{\ell}+\D_{\ell}^{\tps}\meps\D^*_{\ell}\right)^{-1}=\\
&\frac{1}{P_s}\left[ \left(\I_3+\D_{\ell}^{\tps}\meps\D^*_{\ell}\left(\D_{\ell}^{\tps}\D^*_{\ell}\right)^{-1}\right)\left(\D_{\ell}^{\tps}\D^*_{\ell}\right) \right]^{-1}=\\
&\frac{1}{P_s}\left(\D_{\ell}^{\tps}\D^*_{\ell}\right)^{-1}\left(\I_3+\D_{\ell}^{\tps}\meps\D^*_{\ell}\left(\D_{\ell}^{\tps}\D^*_{\ell}\right)^{-1}\right)^{-1}.
\end{aligned}
\end{equation*}
Hence, using the Neumann series \cite{stewart1998matrix}, we have\footnote{By denoting $\mPhi=\mathcal{O}(\meps)$, we mean that $\left|\lambda_{\max}(\mPhi)\right|=\mathcal{O}(\varepsilon_{\max})$, where $\varepsilon_{\max}=|\lambda_{\max}\left(\meps\right)|$. Therefore, $\mPhi\to\mathbf{O}$ when $\varepsilon_{\max}\to0$.}
\begin{gather}
\left(\I_3+\D_{\ell}^{\tps}\meps\D^*_{\ell}\left(\D_{\ell}^{\tps}\D^*_{\ell}\right)^{-1}\right)^{-1}=\I_3+\mathcal{O}(\meps) \; \Longrightarrow\nonumber \\
\left(\D_{\ell}^{\tps}\P_{\sbar}\D^*_{\ell}\right)^{-1}=\frac{1}{P_s}\left[\left(\D_{\ell}^{\tps}\D^*_{\ell}\right)^{-1}+\mathcal{O}(\meps)\right]. \label{approxNeumannseries}
\end{gather}
As expected, the last term in \eqref{approxNeumannseries} indicates that this approximation holds when the deviations from a constant spectral level, quantified here by $\meps$, are sufficiently small with respect to the normalized average power (see \eqref{approxflatsignal}).

Proceeding, by substituting \eqref{approxNeumannseries} into \eqref{Postivecost}, and using well-known eigenvalue perturbation theory results \cite{trefethen1997numerical}, we obtain

\begin{equation*}
\begin{aligned}
&\max_{\text{\boldmath$\bar{s}$}\in\mathcal{S}_N} \bus^{\her}\left(\sum_{\ell=1}^{L}\bX_{\ell}\D^*_{\ell}\left(\D_{\ell}^{\tps}\P_{\sbar}\D^*_{\ell}\right)^{-1}\left(\bX_{\ell}\D^*_{\ell}\right)^{\her}\right)\bus=\\
&\max_{\text{\boldmath$\bar{s}$}\in\mathcal{S}_N} \bus^{\her}\left(\sum_{\ell=1}^{L}\bX_{\ell}\D^*_{\ell}\left[\left(\D_{\ell}^{\tps}\D^*_{\ell}\right)^{-1}\hspace{-0.1cm}+\hspace{-0.05cm}\mathcal{O}(\meps)\right]\left(\bX_{\ell}\D^*_{\ell}\right)^{\her}\right)\bus=\\
&\max_{\text{\boldmath$\bar{s}$}\in\mathcal{S}_N} \bus^{\her}\left[\Q(\up)+\mathcal{O}(\meps)\right]\bus=\lambda_{\max}\left(\Q(\up)\right)+\mathcal{O}(\varepsilon_{\max}),
\end{aligned}
\end{equation*}
where we recall that $\varepsilon_{\max}=|\lambda_{\max}\left(\meps\right)|$ (see \eqref{approxflatsignal}). It follows that
\begin{equation*}
\hup_{\SBL}= \underset{\text{\boldmath$p$}\in\Rset^{3\times1}}{\argmax} \; \lambda_{\max}\left(\Q(\up)\right)+\mathcal{O}(\varepsilon_{\max}).
\end{equation*}
\end{proof}

\section{Proof of Proposition \ref{efficientcomputationmaxeigenval}}\label{proofofprop3}
\begin{proof}
A key observation is that $\Q(\up)$ is low-rank. Indeed, by definition, $\Q(\up)$ is a sum of the following $L$ matrices, 
\begin{equation}\label{defofQlmatrices}
\Q_{\ell}(\up)\triangleq\bX_{\ell}\D^*_{\ell}\left(\D_{\ell}^{\tps}\D^*_{\ell}\right)^{-1}\left(\bX_{\ell}\D^*_{\ell}\right)^{\her}\in\Cset^{N\times N},
\end{equation}
where each is low-rank. Specifically, recall that $\D^{(\ell)}\in\Cset^{N\times3}$, hence 
\begin{equation}\label{firstlowrankmatrix}
\left(\D_{\ell}^{\tps}\D^*_{\ell}\right)^{-1}\in\Cset^{3\times 3} \; \Longrightarrow \; \text{rank}\left(\left(\D_{\ell}^{\tps}\D^*_{\ell}\right)^{-1}\right)=3,
\end{equation}
\addra{where we recall that $\text{rank}\left(\D_{\ell}\right)=3$ by assumption. }In turn, this implies that
\begin{equation*}
\begin{gathered}
\text{rank}\left(\Q_{\ell}(\up)\right)=\text{rank}\left(\bX_{\ell}\D^*_{\ell}\left(\D_{\ell}^{\tps}\D^*_{\ell}\right)^{-1}\left(\bX_{\ell}\D^*_{\ell}\right)^{\her}\right)=3\\
\Longrightarrow \; \text{rank}\left(\Q(\up)\right)=3L,
\end{gathered}
\end{equation*}
assuming $\{\Q_{\ell}(\up)\}$ are linearly independent.\footnote{This holds with probability one, due to the randomness in $\{\bX_{\ell}\}$.} Thus, we conclude that $\Q(\up)$ has only $3L$ non\delra{-}zero eigenvalues. Since typically $L\ll N$, we have established that $\Q(\up)$ is low-rank.

Next, observe that $\Q(\up)$ is a sum of $L$ positive semi-definite matrices, and is therefore a positive semi-definite matrix as well. Due to its special structure \eqref{targetmatrixdef}, it is possible to compute a different matrix, $\widetilde{\Q}(\up)$, with \emph{exactly} the same eigenvalues as those of $\Q(\up)$. For this, define the Cholesky decompositions \cite{golub2013matrix} 
\begin{equation}\label{CholeskyoldAppendix}
\D_{\ell}^{\tps}\D^*_{\ell}\triangleq\mGamma_{\ell}^{\her}\mGamma_{\ell}\in\Cset^{3\times 3}, \; \forall \ell\in\{1,\ldots,L\},
\end{equation}
where $\mGamma_{\ell}\in\Cset^{3\times 3}$. With these $L$ $3$-dimensional square matrices, substituting \eqref{CholeskyoldAppendix} into \eqref{targetmatrixdef}, we may now write
\begin{equation*}
\Q(\up)=\sum_{\ell=1}^{L}\bX_{\ell}\D^*_{\ell}\mGamma_{\ell}^{-1}\left(\bX_{\ell}\D^*_{\ell}\mGamma_{\ell}^{-1}\right)^{\her}\in\Cset^{N\times N},
\end{equation*}
where we emphasize that $\det\left(\mGamma_{\ell}\right)\neq0$ is guaranteed for all $\ell\in\{1,\ldots,L\}$ due to \eqref{firstlowrankmatrix}. Now, define (as in \eqref{Umatrices})
\begin{equation*}
\U(\up)\triangleq\left[\bX_{1}\D^*_{1}\mGamma_{1}^{-1} \cdots \; \bX_{L}\D^*_{L}\mGamma_{L}^{-1}\right]\in\Cset^{N\times 3L},
\end{equation*}
with which
\begin{equation*}
\Q(\up)=\U(\up)\U(\up)^{\her}.
\end{equation*}
However, we have that
\begin{align*}
\Lambda_+\left(\Q(\up)\right)&=\Lambda_+\left(\U(\up)\U(\up)^{\her}\right)\\
&=\Lambda_+\big(\U(\up)^{\her}\U(\up)\big)=\Lambda_+\left(\widetilde{\Q}(\up)\right),
\end{align*}
where $\Lambda_+\left(\C\right)$ denotes the set of the non\delra{-}zero eigenvalues of the semi-positive definite matrix $\C$, and $\widetilde{\Q}(\up)\in\Cset^{3L\times3L}$. Put simply, $\widetilde{\Q}(\up)$ has the same spectrum as $\Q(\up)$. In particular,
\begin{equation*}
\lambda_{\max}\left(\Q(\up)\right)=\lambda_{\max}\left(\widetilde{\Q}(\up)\right).
\end{equation*}
Since $\dim\left(\widetilde{\Q}(\up)\right)=3L< N=\dim\left(\Q(\up)\right)$, we have reduced the computational burden, which is now governed by $L$, rather than $N$. Specifically, the complexity is $\mathcal{O}(NL^2)$, due to the Cholesky decompositions \eqref{CholeskyoldAppendix}, applied to $L$ $3$-dimensional matrices \cite{golub2013matrix}, leading to the matrix multiplication $\U(\up)^{\her}\U(\up)$, and the subsequent application of the power method to the $3L$-dimensional square matrix $\widetilde{\Q}(\up)$. 
\end{proof}

\bibliography{Bibfile}

\begin{thebibliography}{10}
\providecommand{\url}[1]{#1}
\csname url@samestyle\endcsname
\providecommand{\newblock}{\relax}
\providecommand{\bibinfo}[2]{#2}
\providecommand{\BIBentrySTDinterwordspacing}{\spaceskip=0pt\relax}
\providecommand{\BIBentryALTinterwordstretchfactor}{4}
\providecommand{\BIBentryALTinterwordspacing}{\spaceskip=\fontdimen2\font plus
\BIBentryALTinterwordstretchfactor\fontdimen3\font minus
  \fontdimen4\font\relax}
\providecommand{\BIBforeignlanguage}[2]{{%
\expandafter\ifx\csname l@#1\endcsname\relax
\typeout{** WARNING: IEEEtran.bst: No hyphenation pattern has been}%
\typeout{** loaded for the language `#1'. Using the pattern for}%
\typeout{** the default language instead.}%
\else
\language=\csname l@#1\endcsname
\fi
#2}}
\providecommand{\BIBdecl}{\relax}
\BIBdecl

\bibitem{bahr2009cooperative}
A.~Bahr, J.~J. Leonard, and M.~F. Fallon, ``Cooperative localization for
  autonomous underwater vehicles,'' \emph{Int.\ J.\ Robotics Res.}, vol.~28,
  no.~6, pp. 714--728, 2009.

\bibitem{corke2007experiments}
P.~Corke, C.~Detweiler, M.~Dunbabin, M.~Hamilton, D.~Rus, and I.~Vasilescu,
  ``Experiments with underwater robot localization and tracking,'' in
  \emph{Proc.\ IEEE Int.\ Conf.\ Robotics, Automation}, 2007, pp. 4556--4561.

\bibitem{waterston2019ocean}
J.~Waterston, J.~Rhea, S.~Peterson, L.~Bolick, J.~Ayers, and J.~Ellen, ``Ocean
  of things: Affordable maritime sensors with scalable analysis,'' in
  \emph{Prof.\ OCEANS Conf.}, Marseille, France, 2019, pp. 1--6.

\bibitem{bucker1976use}
H.~P. Bucker, ``Use of calculated sound fields and matched-field detection to
  locate sound sources in shallow water,'' \emph{J.\ Acoust.\ Soc.\ Am.},
  vol.~59, no.~2, pp. 368--373, 1976.

\bibitem{tan2011survey}
H.-P. Tan, R.~Diamant, W.~K.~G. Seah, and M.~Waldmeyer, ``A survey of
  techniques and challenges in underwater localization,'' \emph{Ocean
  Engineering}, vol.~38, no. 14-15, pp. 1663--1676, 2011.

\bibitem{chandrasekhar2006localization}
V.~Chandrasekhar, W.~K.~G. Seah, Y.~S. Choo, and H.~V. Ee, ``Localization in
  underwater sensor networks: survey and challenges,'' in \emph{Proc.\ ACM
  Int.\ Workshop Underwater Networks}, 2006, pp. 33--40.

\bibitem{etter1995underwater}
P.~C. Etter, \emph{Underwater acoustic modeling: principles, techniques and
  applications}.\hskip 1em plus 0.5em minus 0.4em\relax CRC {P}ress, 1995.

\bibitem{jensen2011computational}
F.~B. Jensen, W.~A. Kuperman, M.~B. Porter, and H.~Schmidt, \emph{Computational
  Ocean Acoustics}.\hskip 1em plus 0.5em minus 0.4em\relax Springer Science \&
  Business Media, 2011.

\bibitem{chitre2007high}
M.~Chitre, ``A high-frequency warm shallow water acoustic communications
  channel model and measurements,'' \emph{J.\ Acoust.\ Soc.\ Am.}, vol. 122,
  no.~5, pp. 2580--2586, 2007.

\bibitem{cheng2008silent}
X.~Cheng, H.~Shu, Q.~Liang, and D.~H.-C. Du, ``Silent positioning in underwater
  acoustic sensor networks,'' \emph{IEEE Trans. Veh. Technol.}, vol.~57, no.~3,
  pp. 1756--1766, 2008.

\bibitem{kouzoundjian2017tdoa}
B.~Kouzoundjian, F.~Beaubois, S.~Reboul, J.~B. Choquel, and J.-C. Noyer, ``A
  {TDOA} underwater localization approach for shallow water environment,'' in
  \emph{Proc.\ IEEE OCEANS Conf.}, Aberdeen, Scotland, 2017, pp. 1--4.

\bibitem{tuna2017survey}
G.~Tuna and V.~C. Gungor, ``A survey on deployment techniques, localization
  algorithms, and research challenges for underwater acoustic sensor
  networks,'' \emph{Int.\ J.\ Commun.\ Syst.}, vol.~30, no.~17, 2017.

\bibitem{stojanovic1999underwater}
M.~Stojanovic, ``Underwater acoustic communication,'' \emph{Wiley Encyclopedia
  of Electrical and Electronics Engineering}, pp. 1--12, 1999.

\bibitem{diamant2012and}
R.~Diamant, H.-P. Tan, and L.~Lampe, ``{LOS} and {NLOS} classification for
  underwater acoustic localization,'' \emph{IEEE Trans. Mobile Comput.},
  vol.~13, no.~2, pp. 311--323, 2012.

\bibitem{emokpae2011surface}
L.~Emokpae and M.~Younis, ``Surface based anchor-free localization algorithm
  for underwater sensor networks,'' in \emph{Proc.\ IEEE Int.\ Conf.\ Commun.
  (ICC-2011)}, 2011, pp. 1--5.

\bibitem{emokpae2014ureal}
L.~E. Emokpae, S.~DiBenedetto, B.~Potteiger, and M.~Younis, ``{UREAL}:
  {U}nderwater reflection-enabled acoustic-based localization,'' \emph{IEEE
  Sensors J.}, vol.~14, no.~11, pp. 3915--3925, 2014.

\bibitem{baggeroer1993overview}
A.~B. Baggeroer, W.~A. Kuperman, and P.~N. Mikhalevsky, ``An overview of
  matched field methods in ocean acoustics,'' \emph{IEEE J. Ocean. Eng.},
  vol.~18, no.~4, pp. 401--424, 1993.

\bibitem{collins1991focalization}
M.~D. Collins and W.~A. Kuperman, ``Focalization: Environmental focusing and
  source localization,'' \emph{J.\ Acoust.\ Soc.\ Am.}, vol.~90, no.~3, pp.
  1410--1422, 1991.

\bibitem{iscar2017low}
E.~A. Iscar~Ruland, A.~Shree, N.~Goumas, and M.~Johnson-Roberson, ``Low cost
  underwater acoustic localization,'' in \emph{Proc.\ ASA Mtgs.\ Acoust.},
  vol.~30, no.~1, 2017.

\bibitem{gerondeau2020low}
B.~Gerondeau, L.~Galeota, A.~Caudwell, R.~Gouge, A.~Martin, R.~S{\'e}guin, and
  R.~Zitouni, ``Low-cost underwater localization system,'' in \emph{2020
  International Wireless Communications and Mobile Computing (IWCMC)}.\hskip
  1em plus 0.5em minus 0.4em\relax IEEE, 2020, pp. 1153--1158.

\bibitem{mantzel2012compressive}
W.~Mantzel, J.~Romberg, and K.~Sabra, ``Compressive matched-field processing,''
  \emph{J.\ Acoust.\ Soc.\ Am.}, vol. 132, no.~1, pp. 90--102, 2012.

\bibitem{gemba2017adaptive}
K.~L. Gemba, W.~S. Hodgkiss, and P.~Gerstoft, ``Adaptive and compressive
  matched field processing,'' \emph{J.\ Acoust.\ Soc.\ Am.}, vol. 141, no.~1,
  pp. 92--103, 2017.

\bibitem{dubrovinskaya2019bathymetry}
E.~Dubrovinskaya, P.~Casari, and R.~Diamant, ``Bathymetry-aided underwater
  acoustic localization using a single passive receiver,'' \emph{J.\ Acoust.\
  Soc.\ Am.}, vol. 146, no.~6, pp. 4774--4789, 2019.

\bibitem{saucan2020information}
A.~A. Saucan and M.~Z. Win, ``Information-seeking sensor selection for
  ocean-of-things,'' \emph{IEEE Internet of Things J.}, vol.~7, no.~10, pp.
  10\,072--10\,088, 2020.

\bibitem{weiss2004direct}
A.~J. Weiss, ``Direct position determination of narrowband radio frequency
  transmitters,'' \emph{IEEE Signal Process. Lett.}, vol.~11, no.~5, pp.
  513--516, 2004.

\bibitem{knapp1976generalized}
C.~Knapp and G.~Carter, ``The generalized correlation method for estimation of
  time delay,'' \emph{” IEEE Trans. Acoust., Speech, Signal Process.},
  vol.~24, no.~4, pp. 320--327, 1976.

\bibitem{brandstein1997robust}
M.~S. Brandstein and H.~F. Silverman, ``A robust method for speech signal
  time-delay estimation in reverberant rooms,'' in \emph{Proc.\ Int.\ Conf.\
  Acoust., Speech, Signal Process. (ICASSP-1997)}, vol.~1, 1997, pp. 375--378.

\bibitem{grondin2018study}
F.~Grondin and J.~Glass, ``A study of the complexity and accuracy of direction
  of arrival estimation methods based on {GCC}-{PHAT} for a pair of close
  microphones,'' \emph{arXiv preprint arXiv:1811.11787}, 2018.

\bibitem{wang2019robust}
G.~Wang, W.~Zhu, and N.~Ansari, ``Robust {TDOA}-based localization for {I}o{T}
  via joint source position and {NLOS} error estimation,'' \emph{IEEE Internet
  of Things J.}, vol.~6, no.~5, pp. 8529--8541, 2019.

\bibitem{zou2020tdoa}
Y.~Zou and H.~Liu, ``{TDOA} localization with unknown signal propagation speed
  and sensor position errors,'' \emph{IEEE Commun. Lett.}, vol.~24, no.~5, pp.
  1024--1027, 2020.

\bibitem{xiong2021tdoa}
W.~Xiong, C.~Schindelhauer, H.~C. So, J.~Bordoy, A.~Gabbrielli, and J.~Liang,
  ``{TDOA}-based localization with {NLOS} mitigation via robust model
  transformation and neurodynamic optimization,'' \emph{Signal Processing},
  vol. 178, p. 107774, 2021.

\bibitem{hodgkiss2012kauai}
W.~S. Hodgkiss and J.~C. Preisig, ``{Kauai {ACOMMS} {MURI} 2011 (KAM11)
  experiment},'' in \emph{Proc.\ Euro.\ Conf.\ Underwater Acoust.\
  (ECUA-2012)}, 2012, pp. 993--1000.

\bibitem{etter2018underwater}
P.~C. Etter, \emph{Underwater acoustic modeling and simulation}.\hskip 1em plus
  0.5em minus 0.4em\relax CRC press, 2018.

\bibitem{hovem2011ray}
J.~M. Hovem, \emph{Ray trace modeling of underwater sound propagation.
  Documentation and use of the PlaneRay model}, 2011.

\bibitem{cormen2009introduction}
T.~H. Cormen, C.~E. Leiserson, R.~L. Rivest, and C.~Stein, \emph{Introduction
  to Algorithms}.\hskip 1em plus 0.5em minus 0.4em\relax MIT {P}ress, 2009.

\bibitem{aubauer2000one}
R.~Aubauer, M.~O. Lammers, and W.~W.~L. Au, ``One-hydrophone method of
  estimating distance and depth of phonating dolphins in shallow water,''
  \emph{J.\ Acoust.\ Soc.\ Am.}, vol. 107, no.~5, pp. 2744--2749, 2000.

\bibitem{1282692}
F.~Schulz, R.~Weber, A.~Waldhorst, and J.~Bohme, ``Performance enhancement of
  blind adaptive equalizers using environmental knowledge,'' in \emph{Proc.
  OCEANS Conf.}, vol.~4, 2003, pp. 1793--1799.

\bibitem{emokpae2014surface}
L.~Emokpae and M.~Younis, ``Surface-reflection-based communication and
  localization in underwater sensor networks,'' \emph{ACM Trans.\ Sensor
  Networks}, vol.~10, no.~3, pp. 1--51, 2014.

\bibitem{9293153}
A.~Weiss, ``Blind direction-of-arrival estimation in acoustic vector-sensor
  arrays via tensor decomposition and {K}ullback-{L}eibler divergence
  covariance fitting,'' \emph{IEEE Trans. Signal Process.}, vol.~69, pp.
  531--545, 2021.

\bibitem{conn2000trust}
A.~R. Conn, N.~I. Gould, and P.~L. Toint, \emph{Trust Region Methods}.\hskip
  1em plus 0.5em minus 0.4em\relax {SIAM}, 2000.

\bibitem{baggeroer2017stochastic}
A.~B. Baggeroer, ``The stochastic {C}ram\'er-{R}ao bound for source
  localization and medium tomography using vector sensors,'' \emph{J.\ Acoust.\
  Soc.\ Am.}, vol. 141, no.~5, pp. 3430--3449, 2017.

\bibitem{baggeroer1988matched}
A.~B. Baggeroer, W.~Kuperman, and H.~Schmidt, ``Matched field processing:
  Source localization in correlated noise as an optimum parameter estimation
  problem,'' \emph{J.\ Acoust.\ Soc.\ Am.}, vol.~83, no.~2, pp. 571--587, 1988.

\bibitem{collier2005fisher}
S.~L. Collier, ``Fisher information for a complex {G}aussian random variable:
  Beamforming applications for wave propagation in a random medium,''
  \emph{IEEE Trans. Signal Process.}, vol.~53, no.~11, pp. 4236--4248, 2005.

\bibitem{van2014underwater}
W.~A.~P. van Kleunen, K.~C.~H. Blom, N.~Meratnia, A.~B.~J. Kokkeler, P.~J.~M.
  Havinga, and G.~J.~M. Smit, ``Underwater localization by combining
  time-of-flight and direction-of-arrival,'' in \emph{Proc.\ OCEANS Conf.},
  Taipei, Taiwan, 2014, pp. 1--6.

\bibitem{siderius1997multipath}
M.~Siderius, D.~R. Jackson, D.~Rouseff, and R.~Porter, ``Multipath compensation
  in shallow water environments using a virtual receiver,'' \emph{J.\ Acoust.\
  Soc.\ Am.}, vol. 102, no.~6, pp. 3439--3449, 1997.

\bibitem{shulkind2018sensor}
G.~Shulkind, S.~Jegelka, and G.~W. Wornell, ``Sensor array design through
  submodular optimization,'' \emph{IEEE Trans.\ Inf.\ Theory}, vol.~65, no.~1,
  pp. 664--675, 2018.

\bibitem{weiss2021one}
A.~Weiss and G.~W. Wornell, ``One-bit direct position determination of
  narrowband {G}aussian signals,'' in \emph{Proc.\ IEEE Statistical Signal
  Processing Workshop (SSP-2021)}, 2021, pp. 466--470.

\bibitem{stewart1998matrix}
G.~W. Stewart, \emph{Matrix Algorithms: Volume 1: Basic Decompositions}.\hskip
  1em plus 0.5em minus 0.4em\relax {SIAM}, 1998.

\bibitem{trefethen1997numerical}
L.~N. Trefethen and D.~Bau~III, \emph{Numerical Linear Algebra}.\hskip 1em plus
  0.5em minus 0.4em\relax {SIAM}, 1997, vol.~50.

\bibitem{golub2013matrix}
G.~H. Golub and C.~F. Van~Loan, \emph{Matrix Computations}, 3rd~ed.\hskip 1em
  plus 0.5em minus 0.4em\relax Johns Hopkins University {P}ress, 2013.

\end{thebibliography}

\end{document}


\title{A Semi-Blind Method for Localization of Underwater Acoustic Sources \\ \vspace{0.2cm} \smaller{}Supplementary Materials}

\author{Amir Weiss, Toros Arikan, Hari Vishnu, Grant B.\ Deane, Andrew C.\ Singer, and Gregory W. Wornell
}

\maketitle

\section{Derivation of the MFP3 Solution}\label{AppMFP3}
The MFP solution (12) may be simplified, as
\begin{gather}
\sum_{\ell=1}^{L}\left\|\bux_{\ell}-\H_{\ell}\bus\right\|_2^2= \sum_{k=1}^{N}\norm{\bux[k]-\mybar{\uh}_k(\up)\sbar[k]}_2^2\; \label{AppAMFPsimplifiedcost}\\
\Longrightarrow \; \widehat{\sbar}[k]\triangleq \frac{\bux[k]^{\her}\mybar{\uh}_k(\up)}{\|\mybar{\uh}_k(\up)\|_2^2}, \quad \widehat{\bus}\triangleq[\widehat{\sbar}[1] \cdots \widehat{\sbar}[N]]^{\tps},\label{AppAsourceestimateMFP}
\end{gather}
where \eqref{AppAsourceestimateMFP} minimizes \eqref{AppAMFPsimplifiedcost}. Substituting \eqref{AppAsourceestimateMFP} into the cost function $\widetilde{C}_{\MFP}(\up,\bus)$ and further simplifying, we obtain
\begin{align*}
C_{\MFP}(\up)&\triangleq\widetilde{C}_{\MFP}(\up,\widehat{\bus})=\sum_{k=1}^{N}\left\|\bux[k]-\mybar{\uh}_k(\up)\widehat{\sbar}[k]\right\|_2^2\\
&=\underbrace{\sum_{k=1}^{N}\bux[k]^{\her}\bux[k]}_{\substack{\text{constant} \\ {\text{ with respect to \boldmath$p$}}}} -\sum_{k=1}^{N}\bux[k]^{\her}\mybar{\uh}_k(\up)\widehat{\sbar}[k].
\end{align*}
Therefore,
\begin{equation*}
\begin{aligned}
\hspace{-0.1cm}\hup_{\MFP}&=\underset{\text{\boldmath$p$}\in\Rset^{3\times1}}{\argmin} \; C_{\MFP}(\up)=\underset{\text{\boldmath$p$}\in\Rset^{3\times1}}{\argmax} \; \sum_{k=1}^{N}\bux[k]^{\her}\mybar{\uh}_k(\up)\widehat{\sbar}[k]\\
&=\underset{\text{\boldmath$p$}\in\Rset^{3\times1}}{\argmax} \;  \sum_{k=1}^{N}\frac{\left|\bux[k]^{\her}\mybar{\uh}_k(\up)\right|^2}{\|\mybar{\uh}_k(\up)\|^2}.
\end{aligned}
\end{equation*}

\section{Derivation of the FIM Elements}\label{AppACRLB}
Following the discussion in Section VI, a complete description of all the FIM elements is given by (30)--(32). Hence, it remains only to compute the derivatives of $\H\bus$ w.r.t.\ the parameters of $\utheta$, excluding the elements of $\usigma_v^2$ (due to (31)), and to plug in these derivatives in (32) for the computation of all the respective signal-related FIM elements.

We begin with $\uphi_s$, to have
\begin{align}
\frac{\partial\H\bus}{\partial\phi_s[k]}&=\H\left(\frac{\partial\bus}{\partial\phi_s[k]}\right)=\H\left(\sqrt{P_s}\cdot\frac{\partial e^{\jmath\uphi_s}}{\partial\phi_s[k]}\right)\\
&=\H\left(\jmath \sqrt{P_s} e^{\jmath\phi_s[k]}\ue_k\right)=\left(\jmath \sqrt{P_s} e^{\jmath\phi_s[k]}\right)\H\ue_k.
\end{align}
Next, for the real part of the channel attenuation coefficients, using the block structure of $\H$ and the fact that $\H_{\ell}=\Diag(\D_{\ell}\ub_{\ell})$, we have for all $\ell\in\{1,\ldots,L\}$ and $r\in\{1,2,3\}$,
\begin{equation}
\begin{aligned}
\frac{\partial\H\bus}{\partial\Re\{b_{r\ell}\}}&=\left(\frac{\partial\H}{\partial\Re\{b_{r\ell}\}}\right)\bus=\left(\ue_{\ell}\otimes\frac{\partial \Diag(\D_{\ell}\ub_{\ell}) }{\partial\Re\{b_{r\ell}\}}\right)\bus\\
&=\left[\ue_{\ell}\otimes \Diag(\D_{\ell}\ue_{r}) \right]\bus=\ue_{\ell}\otimes \left(\Diag(\D_{\ell}\ue_{r})\bus\right).
\end{aligned}
\end{equation}
Similarly, we have for the imaginary parts,
\begin{equation}
\frac{\partial\H\bus}{\partial\Im\{b_{r\ell}\}}=\jmath\cdot\ue_{\ell}\otimes \left(\Diag(\D_{\ell}\ue_{r})\bus\right).
\end{equation}

We now address the elements of $\up$, which require some additional auxiliary computations. First, observe that
\begin{align}\label{derivative_wrt_x_p}
\frac{\partial\H\bus}{\partial x_p}=\frac{\partial}{\partial x_p}\begin{bmatrix}
\Diag(\D_1\ub_1)\\
\vdots\\
\Diag(\D_L\ub_L)
\end{bmatrix}\bus=\begin{bmatrix}
\Diag(\frac{\partial \text{\boldmath$D$}_1}{\partial x_p}\ub_1)\\
\vdots\\
\Diag(\frac{\partial \text{\boldmath$D$}_L}{\partial x_p}\ub_L)
\end{bmatrix}\bus.
\end{align}
Similar expressions are obtained for $y_p, z_p$ replacing $x_p$ in \eqref{derivative_wrt_x_p}. Hence, it suffices to compute $\left\{\frac{\partial \text{\boldmath$D$}_{\ell}}{\partial x_p}, \frac{\partial \text{\boldmath$D$}_{\ell}}{\partial y_p}, \frac{\partial \text{\boldmath$D$}_{\ell}}{\partial z_p}\right\}$.

The derivative of any coordinate $\mathcal{C}_p\in\{x_p,y_p,z_p\}$ w.r.t.\ the $(k,r)$-th element of $\D_{\ell}$ is given by
\begin{equation}\label{derivative_of_D_wrt_corrdinate}
\frac{\partial [\D_{\ell}]_{kr} }{\partial \mathcal{C}_p}=\frac{\partial e^{\jmath\omega_k\tau_{r\ell}(\text{\boldmath$p$})} }{\partial \mathcal{C}_p}=\jmath\omega_k e^{\jmath\omega_k\tau_{r\ell}(\text{\boldmath$p$})}\cdot\frac{\partial \tau_{r\ell}(\up) }{\partial \mathcal{C}_p}.
\end{equation}
Using the expressions (1)--(3), it can shown with straightforward calculations that for $r\in\{1,2,3\}$,
\begin{equation}\label{derivative_of_tau_x}
\frac{\partial \tau_{r\ell}(\up) }{\partial x_p}=\frac{x_p-x_{\ell}}{c\cdot R_{r\ell}}\triangleq\frac{\Delta x_{\ell}}{c\cdot R_{r\ell}}, \; \forall \ell\in\{1,\ldots,L\}.
\end{equation}
Similarly, from considerations of symmetry, we also have
\begin{equation}
\frac{\partial \tau_{r\ell}(\up) }{\partial y_p}=\frac{y_p-y_{\ell}}{c\cdot R_{r\ell}}\triangleq\frac{\Delta y_{\ell}}{c\cdot R_{r\ell}}, \; \forall \ell\in\{1,\ldots,L\},
\end{equation}
for $r\in\{1,2,3\}$. For the vertical (depth) dimension, with slightly different expressions for the NLOS delays, we have
\begin{align}
&\frac{\partial \tau_{1\ell}(\up) }{\partial z_p}=\frac{z_p-z_{\ell}}{c\cdot R_{1\ell}}\triangleq\frac{\Delta z_{\ell}}{c\cdot R_{1\ell}}, \; \forall \ell\in\{1,\ldots,L\},\\
&\frac{\partial \tau_{2\ell}(\up) }{\partial z_p}=\frac{2z_p}{c\cdot R_{2\ell}}, \quad\quad\quad\quad\;\;\; \forall \ell\in\{1,\ldots,L\}, \\
&\frac{\partial \tau_{3\ell}(\up) }{\partial z_p}=\frac{2(z_p-h)}{c\cdot R_{3\ell}}, \quad\quad\quad\;\; \forall \ell\in\{1,\ldots,L\}.\label{derivative_of_tau_z3}
\end{align}
Finally, by substituting \eqref{derivative_of_tau_x}--\eqref{derivative_of_tau_z3} into \eqref{derivative_of_D_wrt_corrdinate}, and then substituting \eqref{derivative_of_D_wrt_corrdinate} into \eqref{derivative_wrt_x_p}, we obtain all the required derivatives.

\section{Additional Technical Details on the Water Tank Testbed}\label{AppHiFAAT}
The high frequency autonomous acoustic tank provides high frequency ($200$--$400$ kHz) acoustic data from a complex reverberant environment. This apparatus enables us to work directly with experimental data, which inherently contains noise and uncertainty in estimates of geometry and environmental data that may be lacking in simulations, helping to prepare for larger scale experiments.

\subsection{Experimental Configuration}
The water tank testbed is designed around a modified Ender 5 Plus 3D (E5P) printer, which provides precise control of transducer placement and rapid mapping of the tank volume. The E5P has a relatively large print volume of $35\,\text{cm} \times 35\,\text{cm} \times 40\,\text{cm}$ and precise, $0.1\,\text{mm}$ carriage positioning. The tank volume, selected to fit on the Ender 5 plus print table, is roughly $25\,\text{cm} \times 32\,\text{cm} \times 15\,\text{cm}$. Acoustic signals are transmitted and received using ITC 1089D spherical ceramic crystals, which which operate in the $200\,\text{kHz}$--$400\,\text{kHz}$ frequency range. Signals for the ITC1089D transducer are designed using Matlab and uploaded to a networked Siglent SDG2042X arbitrary waveform generator. Received waveforms on the second ITC1089D are amplified and filtered with Stanford Research Systems SIM910 and SIM965 modules before acquisition with an Agilent Technologies DSO5034A oscilloscope. Signal transmission and acquisition is controlled with custom National Instruments Labview software, which also positions the E5P carriage between transmission/acquisition cycles using standard G-code printer commands sent over a virtual serial interface.

The water tank testbed is useful despite its small scale. The tank size is of order of tens of wavelengths in each direction. Experiments have been done with scattering cylinders that are approximately $10$ wavelengths in scale. These experiments demonstrated that for the testbed, the impact of scattering is negligible in the received signal compared to the reverberant arrivals from tank surfaces. The most unrealistic element in the setup is therefore additional reverberation from the tank side-walls. This problem could be addressed but would require re-engineering the carriage mount frame, something that is difficult to do in the current environment. In our localization experiments, the reflections from the tank boundaries are considered as unmodeled phenomena, essentially giving rise to a more difficult scenario (in this respect) than the one the algorithms under consideration would face in open water, where side reflections are typically less probable.

\begin{figure}[t]
	\centering
	{
		\includegraphics[width=0.45\textwidth]{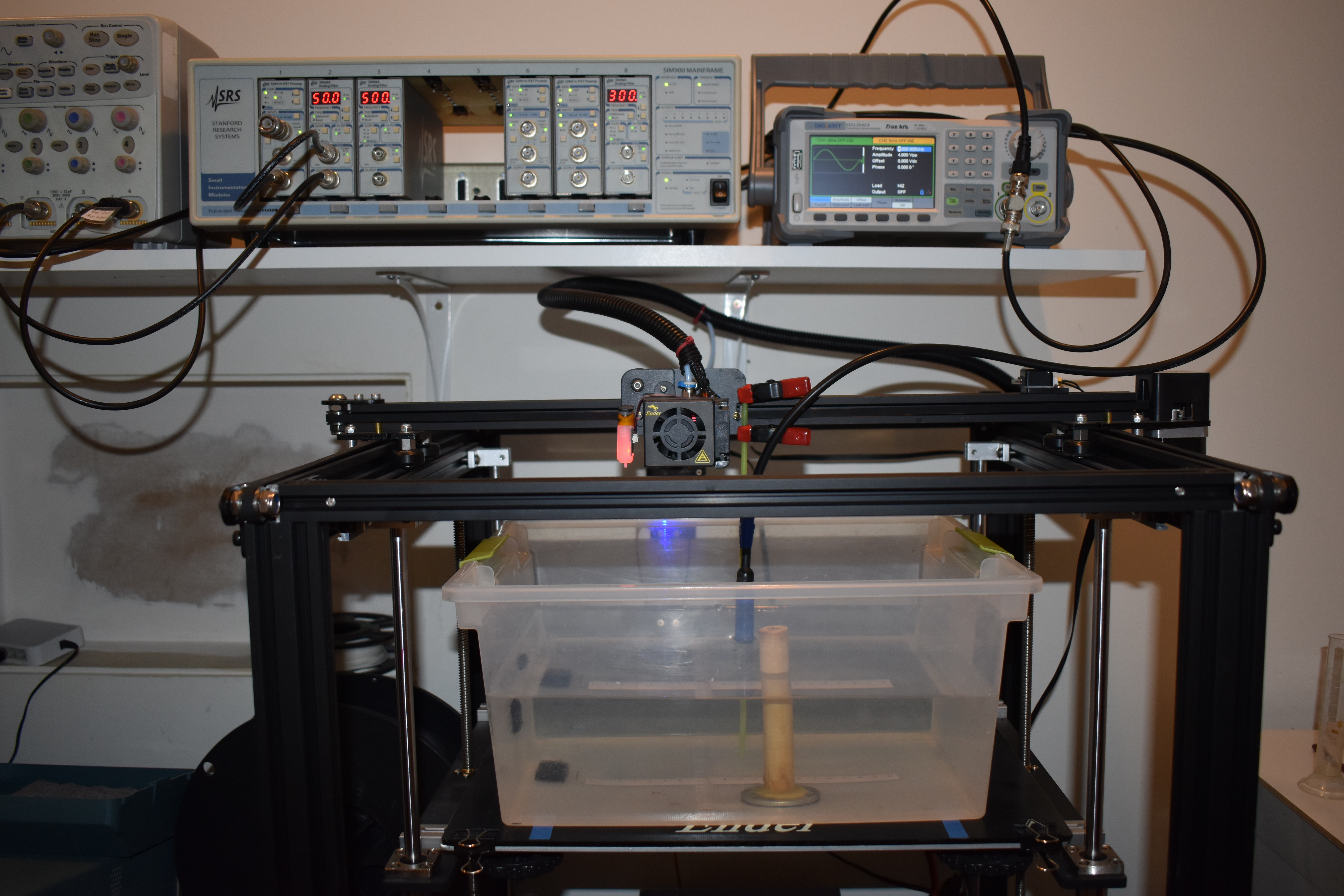}
	}
	\caption{A picture of the water tank showing a ITC1089D transducer mounted near a vertical cylinder. The center frequency of the pulse transmissions is $280$ kHz. The vertical source near the top tank wall is mounted to the carriage of an Ender 5 Plus 3D printer, which provides $3$-axis computer-controlled movement for mapping out tank volumes.}
	\label{fig:HiFAAT}
\end{figure}
\begin{figure}[t]
	\centering
	{
		\includegraphics[width=0.45\textwidth]{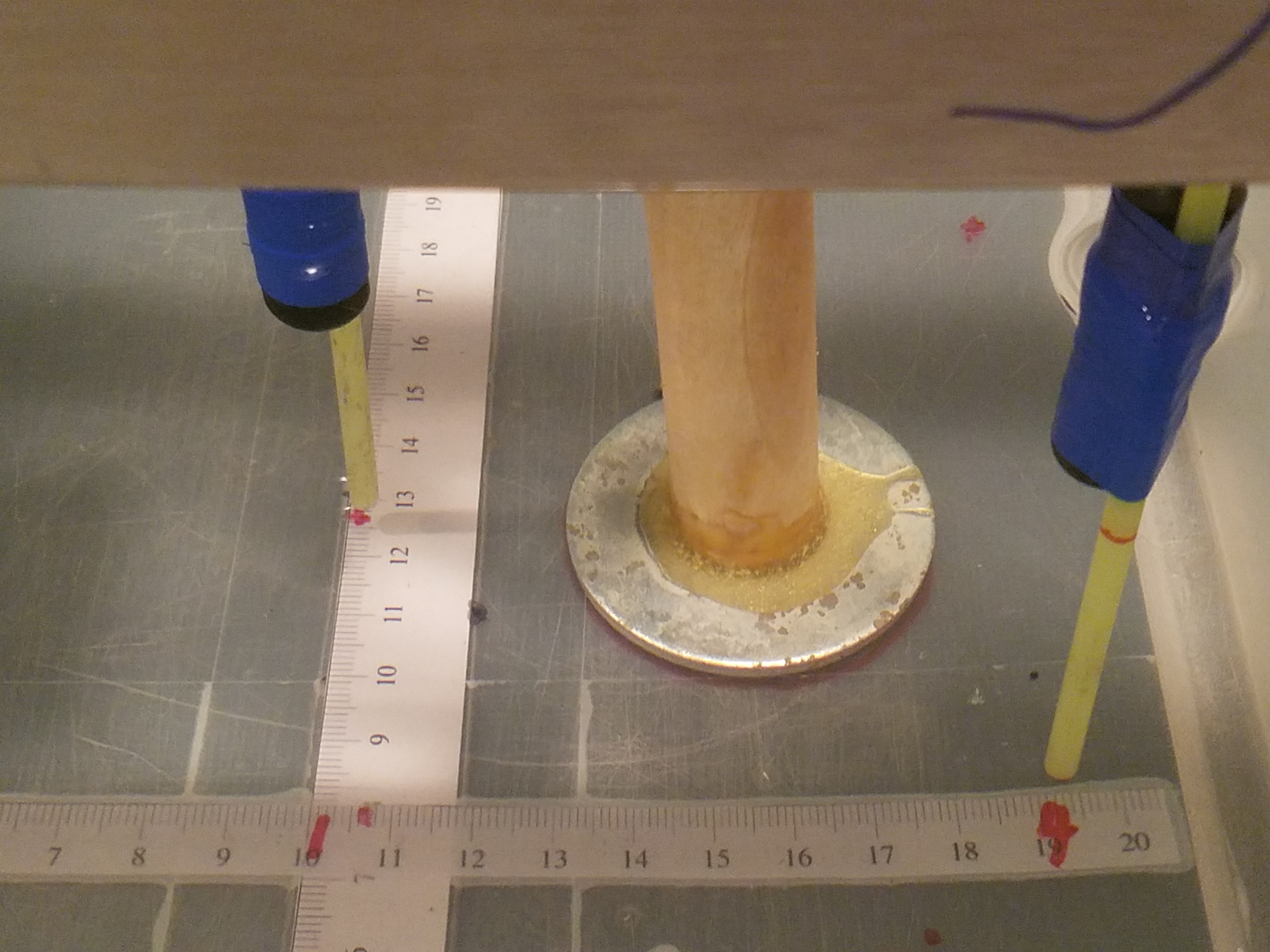}
	}
	\caption{A picture of the water tank testbed, showing the two transducers with the vertical wooden cylinder, modeling an occluder in our experiments.}
	\label{fig:hifaat_setup}
\end{figure}

In our experiments, one of the transducers is used as the source (transmitter) and the other as the receiver. A picture of the setup is given in Fig.\ \ref{fig:hifaat_setup}. We record data for different positions of the source while the receiver is kept stationary. The same transmissions from the same positions of the source are then repeated, with different positions of the receiver, in order to obtain a dataset for an (almost) equivalent setting with multiple receivers (four, in our case). This experiment is conducted twice---once with an occluding cylinder present, and once with it absent.

According to our experiments, the speed of sound in the watertank was $1485\;\text{m}/\text{s}$, which is the value we set for $c$ when applying the localization methods GCC-PHAT, MFP3, and the proposed SBL method. For MFP3, which requires the sediment coefficient as an input parameter as well, our calculations indicate that due to the thinness of the plastic layer at the bottom, the bottom reflection is almost completely returned. Therefore, we set $\kappa_b=1$. Of course, it is most likely that $\kappa_b$ is only approximately $1$, and is slightly smaller. However, in practice it would be virtually impossible to perfectly know the sediment coefficient in the area of interest \textit{a priori}. Hence, here the choice of $\kappa_b=1$ still faithfully represents a scenario in which highly reliable prior environmental knowledge is incorporated into a MFP-based solution.
